\documentclass[12pt]{article}
\usepackage{latexsym}
\usepackage{float}
\usepackage{amsmath,bm,amssymb}
\usepackage{colordvi}
\usepackage{multicol,multirow}
\usepackage{color}
\usepackage{rotating}
 \usepackage{url,hyperref}
\usepackage[ruled]{algorithm2e}
\usepackage{pdflscape}
\usepackage{subfigure}
\usepackage{tabu}
\usepackage{multirow}
\usepackage{booktabs}
\usepackage{graphicx}
\usepackage[utf8]{inputenc}
\usepackage{longtable}

\usepackage[super,sort&compress]{natbib}

\setcitestyle{super}

\def\doublespace{\baselineskip=22pt}
\usepackage{lscape}

\begin{document}
\doublespace
\baselineskip 2.8ex


\begin{center}
{\bf\LARGE Robust Bayesian variable selection for gene-environment interactions}

\vspace{1.6em}

{\bf Jie Ren$^{1,2}$, Fei Zhou$^{2}$, Xiaoxi Li$^{2}$, Shuangge Ma$^{3}$, Yu Jiang$^{4}$ and Cen Wu$^{\ast2}$} \\
\vspace{0.6em}
{ $^1$ Department of Biostatistics, School of Medicine, Indiana University, Indianapolis, IN \\

 $^2$ Department of Statistics, Kansas State University, Manhattan, KS

$^3$ Department of Biostatistics, Yale University, New Haven, CT

$^4$ Division of Epidemiology, Biostatistics and Environmental Health, School of Public Health, University of Memphis, Memphis, TN
}\\

%


\end{center}
%
%
{\bf $\ast$ Corresponding author}:
 Cen Wu, wucen@ksu.edu

\section*{Abstract}

Gene-environment (G$\times$E) interactions have important implications to elucidate the etiology of complex diseases beyond the main genetic and environmental effects. Outliers and data contamination in disease phenotypes of G$\times$E studies have been commonly encountered, leading to the development of a broad spectrum of robust regularization methods. Nevertheless, within the Bayesian framework, the issue has not been taken care of in existing studies. We develop a fully Bayesian robust variable selection method for G$\times$E interaction studies. The proposed Bayesian method can effectively accommodate heavy--tailed errors and outliers in the response variable while conducting variable selection by accounting for structural sparsity. In particular, for the robust sparse group selection, the spike--and--slab priors have been imposed on both individual and group levels to identify important main and interaction effects robustly. An efficient Gibbs sampler has been developed to facilitate fast computation. Extensive simulation studies and analysis of both the diabetes data with SNP measurements from the Nurses’ Health Study and TCGA melanoma data with gene expression measurements demonstrate the superior performance of the proposed method over multiple competing alternatives. \\

\noindent{ Keywords: Bayesian variable selection,  Gene-environment interactions, MCMC, Robust analysis, Sparse group selection} 

\section{Introduction}
\label{makereference4.1}

Deciphering the genetic architecture of complex diseases is a challenging task, as it demands the elucidation of the coordinated function of multiple genetic factors, their interactions, as well as gene-environment interactions. How the genetic contributions to influence the variations in the disease phenotypes are mediated by the environmental factors reveals a unique perspective of the disease etiology beyond the main genetic effects and their interactions (or epistasis) \cite{HUNT,SIMO}. Till now, G$\times$E interaction analyses have been extensively conducted, especially within the framework of genetic association studies\cite{HIR,WLC}, to search for the important main and interaction effects that are associated with the disease trait \cite{MUKH}.

With the availability of a large amount of genetic factors, such as SNPs or gene expressions, G$\times$E interactions are of high dimensionality even though the preselected environmental factors are usually low dimensional. Therefore, variable selection has emerged as a powerful tool to identify G$\times$E interactions associated with the phenotypic traits \cite{FLV,WUM}, and a surging amount of G$\times$E studies have recently been conducted along this line, especially with regularization methods \cite{ZHOU}.

\begin{figure}[h!]
	\centering
	\includegraphics[angle=0,origin=c,width=0.7\textwidth]{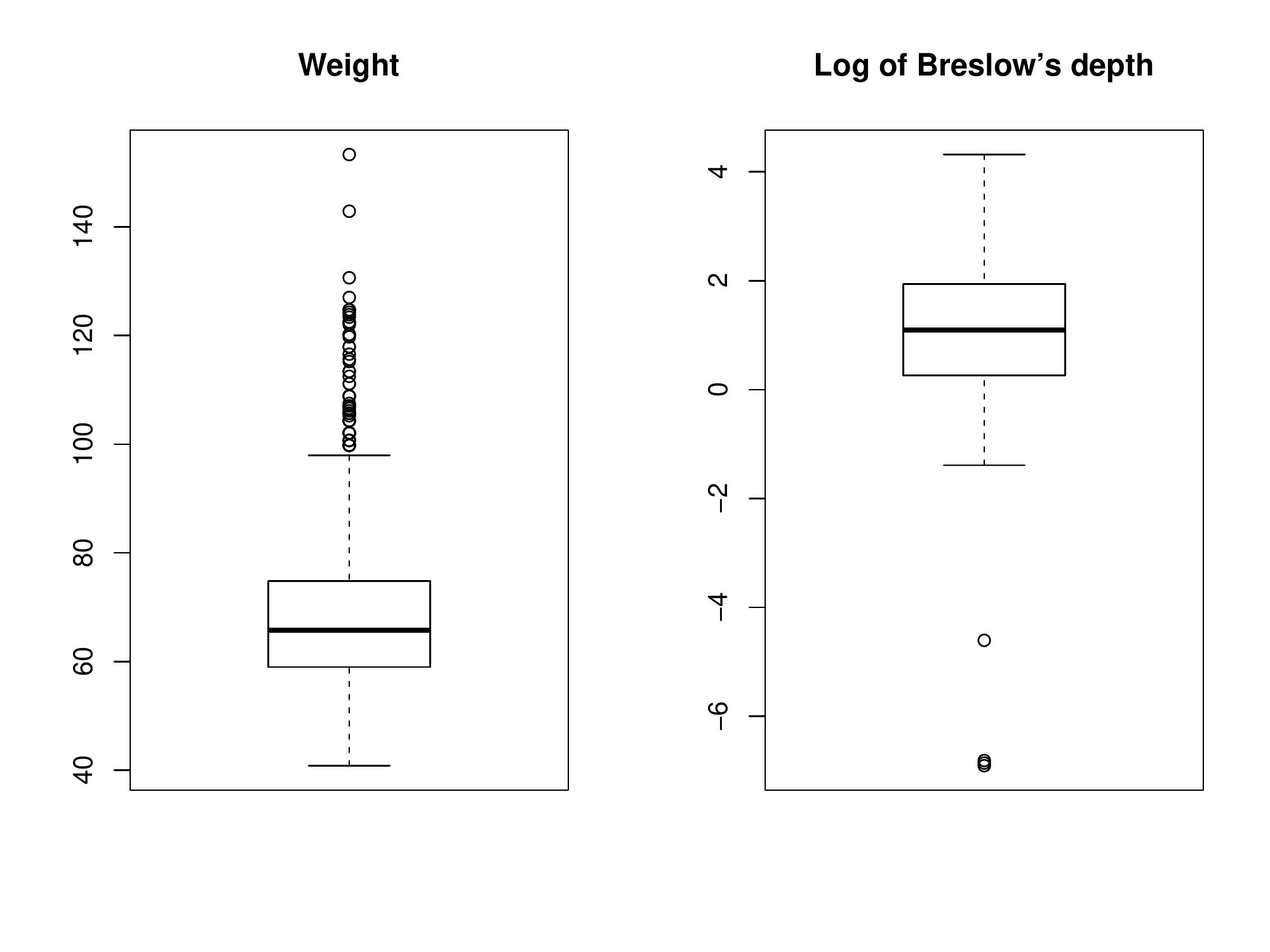}
	\caption[Distribution of the outcome variables]{Distribution of the outcome variables for the NHS (left) and SKCM (right) data.}
	\label{fig:bp}
\end{figure}

A prominent trend among these studies is to incorporate robustness in regularized identification of main and interaction effects in order to accommodate data contamination and heavy-tailed distributions in the disease phenotypes. Take the datasets analyzed in this article for example. The disease outcomes of interest are weight from the Nurses’ Health Study (NHS) and (log-transformed) Breslow’s depth from The Cancer Genome Atlas (TCGA) Skin Cutaneous Melanoma (SKCM) data. We plot the two in Figure \ref{fig:bp}, where the long tails can be clearly observed. In practice, such a heavy-tailed distribution is frequently encountered and arise due to multiple reasons. For instance, some phenotypes have skewness in nature. For the subjects’ recruited for the NHS, their ages are in the range from 41 to 68 as the average age for the onset of type 2 diabetes is 45 \cite{CDC}. The subjects’ weight among this age group does have a right-skewed tendency.  In addition, in the study of complex diseases such as cancer, even patients of similar profiles may have different subtypes as rigorous accrual of patients is usually not affordable. The data from the major disease subtype can be viewed as being “contaminated” by other subtypes or outliers. As nonrobust approaches cannot efficiently accommodate data contamination and long tailed distributions, which inevitably leads to biased estimates and false identifications, the robust regularization methods have thus been extensively developed for G$\times$E studies \cite{WUM,ZHOU}.

Nevertheless, within the Bayesian framework, robust variable selection methods have not been investigated for gene-environment interactions by far. In fact, our literature search indicates that only limited number of Bayesian variable selection methods have been developed for G$\times$E studies, and none of them is robust \cite{ZHOU}. Driven by the urgent need to conduct robust Bayesian analysis, we propose robust Bayesian variable selection methods tailored for interaction studies by adopting a Bayesian formulation of the least absolute deviation (LAD) regression to accommodate data contamination and long-tailed distributions in the phenotype. Such a formulation is a special case of the Bayesian quantile regression \cite{BQUANT}. The LAD loss has been a very popular choice for developing robust regularization methods for data with structured sparsity, including networks \cite{WU2018JMA,REN2019} and sparse group structure \cite{WU2018SM}. Its computational convenience has been revealed within the Bayesian framework as efficient Gibbs sampler can be constructed when the loss is combined with LASSO, group LASSO and elastic net penalties \cite{BRQR}. Furthermore, following the strategy of bi--level selection from a nonrobust Bayesian setting \cite{XXF}, we have developed the Bayesian LAD sparse group LASSO for robust G$\times$E interaction studies. The spike-and-slab priors have been imposed on both the individual and group level to ensure the shrinkage of posterior estimates corresponding to unimportant main and interaction effects to zero exactly. Such a prior leads to the real sparsity and is superior over Laplacian types of shrinkage in terms of identification and prediction results \cite{VSGS,VERO,TANG}.

In this study, our objective is to tackle the challenging task of developing a fully Bayesian robust variable selection method for G$\times$E interactions, which has been well motivated from the success of regularization methods (especially those robust ones) in G$\times$E studies and a lack of robust interaction analysis within the Bayesian framework. The significance of the proposed study lies in the following aspects. First, it advances from existing Bayesian G$\times$E studies by incorporating robustness to accommodate data contamination and heavy-tailed distributions in the disease phenotype. Second, on a broader scope, although robust Bayesian quantile regression based variable selection has been proposed under LASSO, group LASSO and elastic net, the more complicated sparse group (or bi-level) structure, which is of particular importance in high dimensional data analysis in general \cite{BREH}, has not been fully understood yet. We are among the first to develop robust Bayesian sparse group LASSO for bi-level selection. Third, unlike existing Bayesian regularized quantile regression methods which build upon the priors under Laplacian type of shrinkage, we conduct efficient Bayesian regularization on both the individual and group levels by borrowing strength from the spike-and-slab priors, thus leading to better identification and prediction performance over the competing alternatives, as demonstrated in extensive simulation studies and case studies of NHS data with SNP measurements and TCGA melanoma data with gene expression measurements.  To facilitate reproducible research and fast computation using our MCMC algorithms, we implement the proposed and alternative methods in C++, which are available from an open source R package \href{http://CRAN.R-project.org/package=roben}{roben} \cite{roben} on CRAN.

\section{Data and Model Settings}
\label{makereference4.2}

Use subscript $i$ to denote the $i$th subject. Let $(X_{i}, Y_{i}, E_{i}, W_{i})$, ($i=1,\ldots,n$) be independent and identically distributed random vectors. $Y_{i}$ is a continuous response variable representing the phenotypic trait. $X_{i}$ is the $p$--dimensional vector of G factors. The environmental factors and clinical covariates are denoted as the $k$- and $q$-dimensional vectors $E_{i}$ and $W_{i}$, respectively. Considering the following model:


\begin{equation}\label{equr:dataModel}
\begin{aligned}
Y_{i} &= \sum_{t=1}^{q}\alpha_{t}W_{it} + \sum_{m=1}^{k}\theta_{m}E_{im}  + \sum_{j=1}^{p}\gamma_{j}X_{ij} + \sum_{j=1}^{p}\sum_{m=1}^{k}\zeta_{jm}E_{im}X_{ij} +\epsilon_{i} \\
&= \sum_{t=1}^{q}\alpha_{t}W_{it} + \sum_{m=1}^{k}\theta_{m}E_{im}  + \sum_{j=1}^{p} \big(\gamma_{j}X_{ij} + \sum_{m=1}^{k}\zeta_{jm}E_{im}X_{ij}\big) +\epsilon_{i} \\
&= \sum_{t=1}^{q}\alpha_{t}W_{it} + \sum_{m=1}^{k}\theta_{m}E_{im}  + \sum_{j=1}^{p} \big(U_{ij}^\top\beta_{j}\big) +\epsilon_{i},
\end{aligned}
\end{equation}
where $\alpha_{t}$'s, $\theta_{m}$'s, $\gamma_{j}$'s and $\zeta_{jm}$'s are the regression coefficients for the clinical covariates, environmental factors, genetic factors and G$\times$E interactions, respectively. We define $\beta_{j}=(\gamma_{j}, \zeta_{j1},\ldots,\zeta_{jk})^\top \equiv (\beta_{j1},\ldots,\beta_{jL})^\top$ and $U_{ij}=(X_{ij},X_{ij}E_{i1}\ldots,X_{ij}E_{ik})^\top \equiv (U_{ij1},\dots,U_{ijL})^\top$, where $L=k+1$. The coefficient vector $\beta_{j}$ represents all the main and interaction effects corresponding to the $j$th genetic measurement. The $\epsilon_{i}$'s are random errors. Without loss of generality, we assume that the data have been properly normalized so that the intercept can be omitted. Denote $U_{i} = (U_{i1}^\top,\dots, U_{ip}^\top)^\top$, $\alpha=(\alpha_{1},\ldots,\alpha_{q})^\top$, $\theta=(\theta_{1},\ldots,\theta_{k})^\top$ and $\beta=(\beta_{1}^\top,\ldots,\beta_{p}^\top)^\top$. The vector $\beta$ is of length $p\times L$. Then model (\ref{equr:dataModel}) can be written in a more concise form as
\begin{equation}\label{equr:dataModel2}
\begin{aligned}
Y_{i} = W_{i}^\top\alpha + E_{i}^\top\theta + U_{i}^\top\beta +\epsilon_{i}
\end{aligned}
\end{equation}


\subsection{Bayesian LAD Regression}


The least absolute deviation (LAD) regression is well known for its robustness to long-tailed distributions in response. For a Bayesian formulation of LAD regression, we assume that $\epsilon_{i}$'s are i.i.d random variables from the Laplace distribution with density
\begin{equation}\label{equr:error}
f(\epsilon_{i}|\nu) = \frac{\nu}{2}\exp\left\{-\nu |\epsilon_{i}|\right\} \quad i=1,\dots,n,
\end{equation}
where $\nu^{-1}$ is the scale parameter of the Laplace distribution. Let $Y=(Y_{1},\dots,Y_{n})^\top$. With clinical covariates $W=(W_{1},\dots,W_{n})^\top$, environment factors $E=(E_{1},\dots,E_{n})^\top$, and genetic main effects and G$\times$E interactions $U=(U_{1},\dots,U_{n})^\top$, the likelihood function can be expressed as  
\begin{equation}\label{equr:likelihood}
f(Y|W,E,U,\alpha,\theta,\beta,\nu) = \prod_{i=1}^{n}\frac{\nu}{2}\exp\left\{-\nu \left|Y_{i}-\mu_{i}\right|\right\},
\end{equation}
where $\mu_{i} = W_{i}^\top\alpha + E_{i}^\top\theta  + U_{i}^\top\beta$. 

Based on Kozumi and Kobayashi \cite{ALDGS}, the Laplace distribution is equivalent to the mixture of an exponential and a scaled normal distribution. Specifically, let $z$ and $\tilde{u}$ be the standard normal and exponential random variables, respectively. If a random variable ${\epsilon}$ follows the Laplace distribution with parameter $\nu$, then it can be represented as follows
\begin{equation}\label{equr:mix}
{\epsilon}=\nu^{-1}\kappa\sqrt{\tilde{u}}z,
\end{equation}
where $\kappa=\sqrt{8}$ is a constant. Therefore, the response $Y_{i}$ can be rewritten as $Y_{i} = \mu_{i} + \nu^{-1}\kappa\sqrt{\tilde{u_{i}}}z_{i}$, where $z_{i}\thicksim N(0,1)$ and $\tilde{u_{i}}\thicksim \text{Exp}(1)$.
Let $u=\nu^{-1}\tilde{u}$. Then $u$ follows the exponential distribution \text{Exp}($\nu^{-1}$). We thus have the following hierarchical representation of the Laplace likelihood: 
\begin{equation*}
\begin{aligned}
Y_{i} &= \mu_{i} + \nu^{-\frac{1}{2}}\kappa \sqrt{u_{i}}z_{i}, \\
u_{i}|\nu &\stackrel{ind}{\thicksim} \nu\exp\left(-\nu u_{i}\right), \\
z_{i} &\stackrel{ind}{\thicksim} \text{N}(0,\; 1).
\end{aligned}
\end{equation*}
This hierarchical representation allows us to express the likelihood function as a multivariate normal distribution, which is critical to construct a Gibbs sampler for efficient sampling of the regression coefficients corresponding to main and interaction effects robustly.

\textit{Remark}: The Laplace distribution in Bayesian LAD regression can be treated as a special case of the asymmetric Laplace distribution (ALD) in Bayesian quantile regression \cite{ALD,BQUANT}. In Bayesian quantile regression, we assume that $\epsilon_{i}$ follows the asymmetric Laplace distribution with density
\begin{equation}\label{equr:alderror}
f(\epsilon_{i}|\tau,\nu) = \tau(1-\tau)\nu\exp\left\{-\nu\rho_{\tau}(\epsilon_{i})\right\} \quad i=1,\dots,n,
\end{equation}
where the check loss function is $\rho_{\tau}(\epsilon_{i})=\epsilon_{i}\left\{\tau-I(\epsilon_{i}<0)\right\}$ for the $\tau$th quantile ($0<\tau<1$).
Note that, when $\tau=0.5$, the ALD in (\ref{equr:alderror}) reduces to a symmetric Laplace distribution defined in (\ref{equr:error}). Yu and Moyeed \cite{BQUANT} have shown that maximizing a likelihood function under the asymmetric Laplace error distribution (\ref{equr:alderror}) is equivalent to minimizing the check loss function in quantile regression. Kozumi and Kobayashi \cite{ALDGS} have proposed a Gibbs sampler for Bayesian quantile regression based on a location-scale mixture representation of the ALD. Specifically, with $\tilde{u}$ and $z$ defined as above, the asymmetric Laplace error in (\ref{equr:alderror}) can be represented as 
\begin{equation}\label{equr:aldmix}
\epsilon=\nu^{-1}\psi z + \nu^{-1}\kappa\sqrt{\tilde{u}}z,
\end{equation}
where 
\begin{equation*}
\psi = \frac{1-2\tau}{\tau(1-\tau)} \quad \text{and} \quad \kappa = \sqrt{\frac{2}{\tau(1-\tau)}}.
\end{equation*}
When $\tau=0.5$, we have $\psi=0$ and $\kappa=\sqrt{8}$, and equation (\ref{equr:aldmix}) reduces to the Laplace error in (\ref{equr:mix}).

\subsection{Bayesian sparse group variable selection for G$\times$E interactions}
The proposed fully Bayesian sparse group variable selection is motivated by the following considerations. In model (\ref{equr:dataModel}), the coefficient vector $\beta_{j}$ corresponds to the main and interaction effects with respect to the $j$th genetic variant. Whether the genetic variant is associated with the phenotype or not can be determined by whether $\beta_{j}=0$. A zero coefficient vector suggests that the variant does not have any effect on the disease outcome. If $\beta_{j}\neq 0$, then a further investigation on the presence of the main effect, the interaction or both is of interest, which can be facilitated by examining the nonzero component in $\beta_{j}$. Therefore, a tailored robust Bayesian variable selection method for G$\times$E studies should accommodate the selection on both group (the entire vector of $\beta_{j}$) and individual (each component of $\beta_{j}$) levels at the same time. 

In order to impose sparsity on both group and individual level to identify important main and interaction effects, we conduct the decomposition of $\beta_{j}$ by following the reparameterization from \cite{XXF}. Specifically, $\beta_{j}$ is defined as
\begin{equation*}
\beta_{j}=V_{j}^{\frac{1}{2}}b_{j},
\end{equation*}
where $b_{j}=(b_{j1},\dots,b_{jL})^\top$ and $V_{j}^{\frac{1}{2}}=\text{diag}\left\{\omega_{j1},\dots,\omega_{jL} \right\}, \omega_{jl}\ge 0$ ($l=1,...,L$). To determine whether the $j$th genetic variant has any effect at all, we conduct group-level selection on $b_{j}$ by 
adopting the following multivariate spike--and--slab priors
\begin{equation}\label{equr:bj}
\begin{aligned}
b_{j}|\phi^{b}_{j} &\stackrel{ind}{\thicksim} \phi^{b}_{j}\, \text{N}_{L}\left(0,\, \textbf{I}_{L}\right) + (1-\phi^{b}_{j})\delta_{0}(b_{j}), \\
\phi^{b}_{j}|\pi_{0} &\stackrel{ind}{\thicksim} \text{Bernoulli}(\pi_{0}), 
\end{aligned}
\end{equation}
where $\textbf{I}_{L}$ is an identity matrix, $\delta_{0}(b_{j})$ denotes a point mass at $0_{L\times1}$ and $\pi_{0}\in [0,1]$. We introduce a latent binary indicator variable $\phi^{b}_{j}$ for each group $j (j=1,\ldots,p)$ to tackle the group--level selection. In particular, when $\phi^{b}_{j}=0$, the coefficient vector $b_{j}$ has a point mass density at zero and all predictors representing the main and interaction effects in the $j$th group are excluded from the model, indicating that the $j$th genetic factor is not associated with the phenotype. On the other hand, when $\phi^{b}_{j}=1$, the components in coefficient vector $b_{j}$ have non-zero values.

To further determine whether there is an important main genetic effect, G$\times$E interaction or both, we impose sparsity within the group $j$ by assigning the following spike--and--slab priors on each $\omega_{jl} \; (j=1,\ldots,p \;\text{and}\; l=1,\ldots,L)$
\begin{equation}\label{equr:wjl}
\begin{aligned}
\omega_{jl}|\phi^{w}_{jl} &\stackrel{ind}{\thicksim} \phi^{w}_{jl}\, \text{N}^{+}\left(0,\, s^{2}\right) + (1-\phi^{w}_{jl})\delta_{0}(\omega_{jl}), \\
\phi^{w}_{jl}|\pi_{1} &\stackrel{ind}{\thicksim} \text{Bernoulli}(\pi_{1}), 
\end{aligned}
\end{equation}
where $\text{N}^{+}\left(0,\, s^{2}\right)$ denotes a normal distribution, $\text{N}\left(0,\, s^{2}\right)$, truncated below at 0. When the binary indicator variable $\phi^{w}_{jl}=0$, $\omega_{jl}$ is set to zero by the point mass function $\delta_{0}(\omega_{jl})$. Within the $j$th group, when the component $\omega_{jl}=0$, we have $\beta_{jl}=0$ and the corresponding $U_{jl}$ is excluded from the model, even when $b_{j}\neq 0$. This implies that the $j$th genetic variant does not have the main effect (if $l$=1) or the interaction effect with the $(l-1)$th environment factor (if $l>1$). The $\beta_{jl}$ is non-zero if and only if the vector $b_{j}\neq 0$ and the individual element $\omega_{jl}\neq 0$.

In (\ref{equr:bj}) and (\ref{equr:wjl}), $\pi_{0}$ and $\pi_{1}$ control the sparsity on the group and individual level, respectively. Their values should be carefully tuned. Fixing their values at $0.5$ makes the prior essentially non-informative since equal prior probabilities are given to all the sub-models. Instead of fixing $\pi_{0}$ and $\pi_{1}$, we assign conjugate beta priors $\pi_{0} \thicksim \text{Beta}(a_{0}, \, b_{0})$ and $\pi_{1} \thicksim \text{Beta}(a_{1}, \, b_{1})$, which can automatically account for the uncertainty in choosing $\pi_{0}$ and $\pi_{1}$. We fixed parameters $a_{0}=b_{0}=a_{1}=b_{1}=1$, so that the priors are essentially non-informative. For computation convenience, we assign a conjugate Inverse--Gamma hyperprior on $s^{2}$
\begin{equation*}
s^{2} \thicksim \text{Inv-Gamma}(1, \eta)
\end{equation*}
$\eta$ is estimated with the Monte Carlo EM algorithm \cite{PARK,XXF}. For the $g$th EM update,
\begin{equation*}
\eta^{(g)} = \frac{1}{E_{\eta^{(g-1)}} \left[\frac{1}{s^{2}}|Y\right]},
\end{equation*}
where the posterior expectation of $\frac{1}{s^{2}}$ is estimated from the MCMC samples based on $t^{(g-1)}$. To maintain conjugacy, we place a Gamma prior on $\nu$,
\begin{equation*}
\nu \thicksim \text{Gamma}(c, d).
\end{equation*}
where $c$ and $d$ are set to small values.
\subsection{Gibbs sampler}
The joint posterior distribution of all the unknown parameters conditional on data can be expressed as
\begin{equation*}\label{equr:full}
\begin{aligned}
\pi (\alpha, \theta, b_{j}, \omega_{jl}, & \nu, u_{i}, \pi_{0}, \pi_{1}, s^{2}|Y) \\
\propto & \prod_{i=1}^{n}(2\pi\kappa^{2}\nu^{-1}u_{i})^{-\frac{1}{2}} \exp\left\{-\frac{\left(Y_{i} - W_{i}^\top\alpha - E_{i}^\top\theta  - \sum_{j=1}^{p} \big(U_{ij}^\top\beta_{j}\big)\right)^{2}}{2\kappa^{2}\nu^{-1}u_{i}}\right\} \\
&\times\prod_{i=1}^{n}\nu\exp(-\nu u_{i}) \; \nu^{c-1}\exp\left\{-d\nu\right\}\\
&\times \exp\Big(-\frac{1}{2}\theta^\top \Sigma_{\theta0}^{-1}\theta\Big) \exp\Big(-\frac{1}{2}\alpha^\top \Sigma_{\alpha0}^{-1}\alpha\Big)  \\
& \times \prod_{j=1}^{p}\left( \pi_{0}(2\pi)^{-\frac{L}{2}} \exp\left\{ -\frac{1}{2}b_{j}^\top b_{j}\right\} \textbf{I}_{\{b_{j} \neq 0\}}+ (1-\pi_{0})\delta_{0}(b_{j}) \right) \\
& \times \prod_{j=1}^{p}\prod_{l=1}^{L} \left( \pi_{1}2(2\pi s^{2})^{-\frac{1}{2}} \exp\left\{ -\frac{\omega_{jl}^{2}}{2s^{2}}\right\} \textbf{I}_{\{\omega_{jl} > 0\}}+ (1-\pi_{1})\delta_{0}(\omega_{jl}) \right) \\
& \times \pi_{0}^{a_{0}-1}(1-\pi_{0})^{b_{0}-1} \\
& \times \pi_{1}^{a_{1}-1}(1-\pi_{1})^{b_{1}-1} \\
& \times (s^{2})^{-2}\exp(-\frac{\eta}{s^{2}}).
\end{aligned}
\end{equation*}
Define the coefficient vector without the $j$th group as $\beta_{(j)}=(\beta_{1}^\top,\dots,\beta_{j-1}^\top, \beta_{j+1}^\top, \dots, \beta_{p}^\top)$ and the corresponding part of the design matrix as $U_{(j)}$. Likewise, define the coefficient vector without the $l$th element in the $j$th group as $\beta_{(jl)}$ and the corresponding design matrix as $U_{(jl)}$. Let $l^{b}_{j}= p(b_{j} \neq 0 | \text{rest})$, then the conditional posterior distribution of $b_{j}$ is a multivariate spike--and--slab distribution:
\begin{equation}\label{equr:prstar}
b_{j}|\text{rest} \thicksim l^{b}_{j}\, \text{N}_{L}(\mu_{b_{j}}, \, \Sigma_{b_{j}}) + (1-l^{b}_{j})\, \delta_{0}(b_{j}),
\end{equation}
where  $\Sigma_{b_{j}} = \left(\nu\kappa^{-2}\sum_{i=1}^{n}u_{i}^{-1}V_{j}^{\frac{1}{2}}U_{ij}U_{ij}^\top V_{j}^{\frac{1}{2}} + \textbf{I}_{L}\right)^{-1}$,  $\mu_{b_{j}}=\Sigma_{b_{j}}\nu\kappa^{-2}\sum_{i=1}^{n}u_{i}^{-1}V_{j}^{\frac{1}{2}}U_{ij}\tilde{y}_{ij}$ and $\tilde{y}_{ij}=y_{i}-W_{i}^\top\alpha-E_{i}^\top\theta-U_{(j)}^\top\beta_{(j)}$. The $l^{b}_{j}$ can be derived as
\begin{equation*}
\setlength{\jot}{10pt}
\begin{aligned}
l^{b}_{j} = \frac{\pi_{0}}{\pi_{0} + (1-\pi_{0})|\Sigma_{b_{j}}|^{-\frac{1}{2}} \exp\left\{- \frac{1}{2} \lVert\Sigma_{b_{j}}^{\frac{1}{2}}\nu\kappa^{-2}\sum_{i=1}^{n}u_{i}^{-1}V_{j}^{\frac{1}{2}}U_{ij}\tilde{y}_{ij}\rVert_{2}^{2} \right\}}. 
\end{aligned}
\end{equation*}
The posterior distribution (\ref{equr:prstar}) is a mixture of a multivariate normal and a point mass at $0$. Specifically, at the $g$th iteration of MCMC, $b_{j}^{(g)}$ is drawn from $N(\mu_{b_{j}}, \, \Sigma_{b_{j}})$ with probability $l_{j}^{b}$ and is set to $0$ with probability $1-l_{j}^{b}$. If $b_{j}^{(g)}$ is set to $0$, we have $\phi_{j}^{b(g)}=0$, which suggests that the $j$th genetic variant is not associated with the phenotype at the $g$th iteration. Otherwise, $\phi_{j}^{b(g)}=1$.

In addition to the multivariate spike--and--slab distribution on the group level, on the individual level, the conditional posterior distribution of $\omega_{jl}$ is also spike-and-slab. Let $l^{w}_{jl}= p(\omega_{jl} \neq 0 | \text{rest})$, we have 
\begin{equation*}
\omega_{jl}|\text{rest} \thicksim l^{w}_{jl}\; \text{N}^{+}(\mu_{\omega_{jl}}, \, \sigma^{2}_{\omega_{jl}}) + (1-l^{w}_{jl})\delta_{0}(\omega_{jl}),
\end{equation*}
where $\sigma^{2}_{\omega_{jl}} = \left(\frac{1}{s^{2}} + \nu\kappa^{-2}\sum_{i=1}^{n}u_{i}^{-1}U_{ijl}^{2} b_{jl}^{2}\right)^{-1}$, $\mu_{\omega_{jl}}=\sigma_{\omega_{jl}}^{2}\nu\kappa^{-2}\sum_{i=1}^{n}u_{i}^{-1}b_{jl}U_{ijl}\tilde{y}_{ijl}$ and $\tilde{y}_{ijl}=y_{i}-W_{i}^\top\alpha-E_{i}^\top\theta-U_{(jl)}^\top\beta_{(jl)}$. It can be shown that
\begin{equation*}
\begin{aligned}
l^{w}_{jl} = \frac{\pi_{1}}{\pi_{1} + (1-\pi_{1})\frac{1}{2}s(\sigma_{\omega_{jl}}^{2})^{-\frac{1}{2}} \exp\left\{- \frac{1}{2}\sigma_{\omega_{jl}}^{2} \left(\nu\kappa^{-2}\sum_{i=1}^{n}u_{i}^{-1}b_{jl}U_{ijl}\tilde{y}_{ijl}\right)^{2} \right\}\left[\Phi\left(\frac{\mu_{\omega_{jl}}}{\sigma_{\omega_{jl}}}\right)\right]^{-1}}, 
\end{aligned}
\end{equation*}
where $\Phi(\cdot)$ is the cumulative distribution function of the standard normal random variable. At the $g$th iteration, the value of $\phi_{jl}^{w(g)}$ can be determined by whether the $\omega_{jl}^{(g)}$ is set to $0$ or not. Recall that $\phi_{jl}^{w(g)}=0$ implies that the $j$th genetic variant does not have the main effect (if $l$=1) or  the interaction effect with the $(l-1)$th E factor (if $l>1$).

The full conditional distribution for $u_{i}$ is Inverse-Gaussian:
\begin{equation*}
u_{i}|\text{rest} \thicksim \text{Inverse-Gaussian}(\mu_{u_{i}}, \, \lambda_{u_{i}}),
\end{equation*}
where the shape parameter $\lambda_{u_{i}}=2\nu$, mean parameter $\mu_{u_{i}}=\sqrt{\frac{2\kappa^{2}}{(Y_{i}-\tilde{y_{i}})^{2}}}$ and $\tilde{y_{i}}=Y_{i}-W_{i}^\top\alpha-E_{i}^\top\theta-U_{i}^\top\beta$.

With the conjugate Inverse--Gamma prior, the posteriors of $s^{2}$ is still an Inverse--Gamma distribution
\begin{equation*}
s^{2}|\text{rest} \thicksim \text{Inv-Gamma}\left(1+\frac{1}{2}\sum_{j,l}\textbf{I}_{\{\omega_{jl} \neq 0\}},\; \eta+\frac{1}{2}\sum_{j,l}\omega_{jl}^{2}\right).
\end{equation*}

With conjugate Beta priors, $\pi_{0}$ and $\pi_{1}$ have beta posterior distributions
\begin{equation*}\label{equr:pphi}
\setlength{\jot}{3pt}
\begin{aligned}
\pi_{0}|\text{rest} &\thicksim \text{Beta}\left(a_{0}+\sum_{j=1}^{p}\textbf{I}_{\{b_{j} \neq 0\}}, \; b_{0}+\sum_{j=1}^{p}\textbf{I}_{\{b_{j} = 0\}}\right), \\
\pi_{1}|\text{rest} &\thicksim \text{Beta}\left(a_{1}+\sum_{j,l}\textbf{I}_{\{\omega_{jl} \neq 0\}}, \; b_{1}+\sum_{j,l}\textbf{I}_{\{\omega_{jl} = 0\}}\right).
\end{aligned}
\end{equation*}
Last, the full conditional distribution for $\nu$ is Gamma distribution
\begin{equation*}
\nu|\text{rest} \thicksim \text{Gamma}\left(s_{\nu}, \, r_{\nu} \right),
\end{equation*}
where the shape parameter $s_{\nu}=c+\frac{3n}{2}$ and the rate parameter $r_{\nu}=d+\sum_{i=1}^{n}u_{i}+(2\kappa^{2})^{-1}\sum_{i=1}^{n}u_{i}^{-1}\tilde{y_{i}}^{2}$. 
Under our prior setting, conditional posterior distributions of all unknown parameters have closed forms by conjugacy. Therefore, efficient Gibbs sampler can be constructed for the posterior distribution. 

We term the proposed robust Bayesian sparse group variable selection with spike and slab priors as RBSG--SS, with direct competitors RBG--SS, RBL--SS and ones without spike and slab priors: RBSG, RBG and RBL. With the non-robust counterpart, there are 12 methods under comparison, which have all been implemented in the C++ based R package \href{http://CRAN.R-project.org/package=roben}{roben} \cite{roben} available from CRAN. It is worth mentioning that besides RBSG--SS, RBG--SS, RBL--SS and RBSG have also been proposed for the first time. A summary of all the methods is provided below.

\subsection{A summary of proposed and alternative methods}
All the methods under comparison can be grouped according to three criteria: with or without robustness, with or without spike-and-slab priors, and the types of structured sparsity (individual-, group- and bi-level) accommodated through variable selection. We first describe the robust Bayesian methods with spike-and-slab priors: RBSG--SS, RBG--SS and RBL--SS, which have all been proposed for the first time. Among them, RBSG--SS is the “golden” method developed for conducting robust sparse group variable selection for G$\times$E interactions with spike-and-slab priors on both the group and individual levels. Besides, RBG--SS and RBL--SS are robust Bayesian group level and individual level selection with spike--and--slab priors, respectively. The spike--and--slab prior has only been imposed on the group level in RBG--SS. Compared to RBSG--SS, it does not induce within group sparsity. On the other hand, RBL--SS conducts individual-level selection without accounting for group structure. An immediate family of robust methods related to the three are RBSG, RBG and RBL, which do not adopt spike--and--slab priors and cannot shrink coefficients corresponding to the main and interaction effects to zero exactly. While RBG and RBL can be directly derived based on Li et al. \cite{BRQR}, RBSG, robust Bayesian sparse group selection, has not been investigated in existing studies so far.

We have also included six non-robust methods for comparison. Among them, BSG--SS, BG--SS and BL--SS are the non--robust counterparts of RBSG--SS, RBG--SS and RBL--SS, respectively. In particular, the BSG--SS conducts (non--robust) Bayesian sparse group selection with spike-and-slab priors on group and individual level simultaneously, while variable selection has only been conducted on group (individual) level through RBG--SS (RBL--SS) under the spike--and--slab priors. In addition, BSG, BG and BL can be viewed as the benchmarks without incorporating spike--and--slab priors corresponding to BSG--SS, BG--SS and BL--SS. They can also be considered as the non--robust counterpart corresponding to RBSG, RBG and RBL. All the six non--robust alternatives can be readily derived based on existing studies.

For clarification, we list all the methods under comparison in Table \ref{methods} in the Appendix. Our contribution includes developing the 4 robust Bayesian variable selection approaches, RBSG--SS, RBG--SS, RBL--SS and RBSG among the first time. For all the rest of the approaches, a modification to the methods from the references provided in Table \ref{methods} by including clinical covariates is necessary. Otherwise, these methods cannot be adopted for a direct comparison with the four newly developed ones.

\section{Simulation}
\label{makereference4.3}

We comprehensively evaluate the proposed and alternative methods through simulation studies. Under all the settings, the responses are generated from model (\ref{equr:dataModel}) with $n=500, q=3, p=100$ and $k=5$, which leads to a total dimension of 608 with 105 main effects, 500 interactions and 3 additional clinical covariates. The genetic main effects and G$\times$E interactions form 100 groups with group size $L=6$. We consider six error distributions for  $\epsilon_{i}$'s: N(0, 1)(Error 1), Laplace($\mu$,$b$) with the mean $\mu=0$ and the scale parameter $b=2$ (Error 2), 10\%Laplace(0,1) + 90\%Laplace(0,$\sqrt{5}$) (Error 3), 90\%N(0,1) + 10\%Cauchy(0, 1) (Error 4), t-distribution with 2 degrees of freedom ($t(2)$) (Error 4), LogNormal(0,1) (Error 5). All of them are heavy-tailed distributions except the first one. 


We assess the performance in terms of identification and prediction accuracy. For methods incorporating spike--and--slab priors, we consider the median probability model (MPM) \cite{XXF,BAR} to identify important effects. In particular, for the proposed RBSG--SS, we define $\phi_{jl}=\phi^{b}_{j}\phi^{w}_{jl}$ for the $l$th predictor in the $j$th group. At the $g$th MCMC iterations, this predictor is included in the model if the indicator $\phi_{jl}^{(g)}$ is 1. Suppose we have collected G posterior samples from the MCMC after burn-ins, then the posterior probability of including the $l$th predictor from the $j$th group in the final model is
\begin{equation}\label{equr:pj}
p_{jl}=\hat{\pi}(\phi_{jl}=1|y)=\frac{1}{G}\sum_{g=1}^{G} \phi_{jl}^{(g)}, \quad j=1,\dots, p \;\text{and}\; l=1,\dots, L.
\end{equation}
A higher posterior inclusion probability $p_{jl}$ can be interpreted as a stronger empirical evidence that the corresponding predictor has a non-zero coefficient and is associated with the phenotype. The MPM model is defined as the model consisting of predictors with at least $\frac{1}{2}$ posterior inclusion probability. When the goal is to select a single model, Barbieri and Berger \cite{BAR} recommend using MPM  because of its optimal prediction performance. Meanwhile, the 95\% credible interval (95\%CI) \cite{LJH} is adopted for methods without spike--and--slab priors.  

Prediction performance is evaluated using the mean prediction errors on an independently generated testing dataset under the same data generating model over 100 replicates. For all robust approaches, the prediction error is defined as mean absolute deviations (MAD). MAD can be computed as $\frac{1}{n}\sum_{i=1}^{n}\left|y_{i}-\hat{y_{i}}\right|$. The prediction error for non--robust ones is defined as the mean squared error (MSE), i.e.,  $\frac{1}{n}\sum_{i=1}^{n}\left(y_{i}-\hat{y_{i}}\right)^{2}$.

The G factors are simulated in the following 4 examples(settings). In the first example, a gene expression matrix with $n=500$ and $p=100$ has been generated from a multivariate normal distribution with marginal mean 0, marginal variance 1 and an auto-regression correlation structure ($\rho=0.3$). In the second example,  the single-nucleotide polymorphism (SNP) data are obtained by dichotomizing the gene expression values (from the first setting) at the 1st and 3rd quartiles, with the 3--level (0,1,2) for genotypes (aa,Aa,AA) respectively. In the third setting, the SNP data are simulated under a pairwise linkage disequilibrium (LD) structure. Let the minor allele frequencies (MAFs) of two neighboring SNPs with risk alleles A and B be $r_{1}$ and $r_{2}$, respectively. The frequencies of four haplotypes are as $p_{AB} = r_{1}r_{2} + \delta$, $p_{ab} = (1-r_{1})(1-r_{2})+\delta$, $p_{Ab} = r_{1}(1-r_{2})-\delta$, and $p_{aB} = (1-r_{1})r_{2}-\delta$, where $\delta$ denotes the LD. Assuming Hardy-Weinberg equilibrium and given the allele frequency for A at locus 1, we can generate the SNP genotype (AA, Aa, aa) from a multinomial distribution with frequencies $(r^{2}_{1}, 2r_{1}(1-r_{1}),(1-r_{1})^{2})$. The genotypes at locus 2 can be simulated according to the conditional genotype probability matrix in Cui et al. \cite{CYH}. We have $\delta=r_{p}\sqrt{r_{1}(1-r_{1})r_{2}(1-r_{2})}$ with MAFs 0.3 and pairwise correlation $r_{p} = 0.6$. In the last example, we consider a more practical correlation structure by extracting the first 100 SNPs from the NHS data analyzed in the case study, so the correlation is based on the real data. For each simulation replicate, we randomly sample 500 subjects from the dataset. 

For E factors, five continuous variables are generated from a multivariate normal distribution with marginal mean 0, marginal variance 1 and AR correlation structure with $\rho=0.5$. We then dichotomize one of them at 0 to create a binary E factor. Besides, we simulate three clinical covariates 
from a multivariate normal distribution and AR correlation structure with $\rho=0.5$, and dichotomize one of them at 0 to create a binary variable.

For the clinical covariates and environmental main effects, the coefficients $\alpha_{t}$'s and $\theta_{m}$'s are generated from Uniform$[0.8, 1.5]$. For genetic main effect and G$\times$E interactions, we randomly selected 25 $\beta_{jl}$'s in 9 groups to have non-zero values that are generated from Uniform$[0.3, 0.9]$. All other $\beta_{jl}$'s are set to zeros.

\begin{table} [h!]
	\def\arraystretch{1.5}
	\begin{center}
		\caption[Simulation results in Example 1.]{Simulation results in Example 1. $(n,q,k,p)$ = (500, 2, 5, 100). mean(sd) of true positives (TP), false positives (FP) and prediction errors (Pred) based on 100 replicates.}\label{perf.gene.ss}
		\centering
		\fontsize{10}{12}\selectfont{
			\begin{tabu} to \textwidth{ X[0.8c] X[0.5r] X[c] X[c] X[c] | X[c] X[c] X[c]}
				\hline
				&&  RBSG--SS  & RBG--SS & RBL--SS  & BSG--SS    & BG--SS & BL--SS      \\
				\cmidrule(lr){1-8}
				\textbf{Error 1} &TP &24.97(0.18)	&25.00(0.00)	&24.93(0.25)	&24.97(0.18)	&25.00(0.00)	&24.93(0.25)
				\\
				{}&FP	&1.30(1.24)	&29.60(2.42)	&1.30(1.44)	&0.47(0.68)	&29.00(0.00)	&0.43(0.73)
				\\
				&Pred	&0.83(0.03)	&0.86(0.03)	&0.84(0.04)	&1.07(0.07)	&1.13(0.07)	&1.08(0.08)
				\\ \cmidrule(lr){1-8}
				\textbf{Error 2}  &TP	&21.66(1.72)	&24.84(0.55)	&18.58(2.14)	&19.98(1.95)	&24.58(0.86)	&15.54(2.04)	
				\\
				{}&FP	&1.32(1.33)	&30.96(4.27)	&1.62(1.64)	&1.82(1.53)	&30.98(4.83)	&0.92(0.94)
				\\
				&Pred &2.15(0.10)	&2.17(0.09)	&2.24(0.12)	&9.32(0.97)	&8.98(0.79)	&10.09(1.08)
				\\ \hline
				\textbf{Error 3} &TP &21.28(2.24)	&24.80(0.73)	&18.14(2.68)	&19.00(2.61)	&24.40(1.09)	&14.24(2.39)
				\\
				{}&FP	&1.48(1.34)	&30.64(4.23)	&1.42(1.63)	&2.04(1.73)	&30.20(4.73)	&1.18(1.16)
				\\
				&Pred &2.29(0.12)	&2.32(0.11)	&2.41(0.12)	&11.11(1.12)	&10.59(0.95)	&12.02(1.12)
				\\ \hline
				\textbf{Error 4} &TP &23.80(1.30)	&24.93(0.37)	&21.80(1.94)	&16.20(6.45)	&21.83(5.24)	&12.53(5.79)		
				\\
				{}&FP &0.53(0.86)	&29.47(2.56)	&0.20(0.41)	&3.73(4.61)	&35.77(23.92)	&1.93(2.49)
				\\
				&Pred &1.50(0.14)	&1.52(0.13)	&1.53(0.14)	&12.48(6.56)	&12.34(7.27)	&13.35(6.72)
				\\
				\hline
				\textbf{Error 5} &TP &24.33(0.76)	&25.00(0.00)	&22.93(1.20)	&22.93(1.26)	&25.00(0.00)	&18.00(2.17)
				\\
				{}&FP &0.26(0.45)	&29.00(0.00)	&0.13(0.35)	&4.30(3.40)	&34.80(8.11)	&1.23(1.55)
				\\
				&Pred &1.16(0.10)	&1.18(0.10)	&1.18(0.10)	&4.75(1.24)	&4.78(1.23)	&5.18(1.34)
				\\
				\hline
		\end{tabu} }
	\end{center}
	\centering
\end{table}

\begin{table} [h!]
	\def\arraystretch{1.5}
	\begin{center}
		\caption[Simulation results in Example 1 (2).]{Simulation results in Example 1. $(n,q,k,p)$ = (500, 2, 5, 100). mean(sd) of true positives (TP), false positives (FP) and prediction errors (Pred) based on 100 replicates.}\label{perf.gene}
		\centering
		\fontsize{10}{12}\selectfont{
			\begin{tabu} to \textwidth{ X[0.8c] X[0.5r] X[c] X[c] X[c] | X[c] X[c] X[c]}
				\hline
				&&  RBSG  & RBG & RBL  & BSG    & BG & BL      \\
				\cmidrule(lr){3-8}
				\textbf{Error 1} &TP &21.87(1.38)	&24.67(0.76)	&21.97(1.40)	&22.93(1.34)	&24.93(0.37)	&23.07(1.23)
				\\
				&FP	&2.63(1.94)	&55.33(15.76)	&3.07(2.35)	&2.43(1.77)	&83.47(20.07)	&11.20(4.34)
				\\
				&Pred	&1.15(0.05)	&1.37(0.06)	&1.15(0.05)	&1.73(0.12)	&2.29(0.19)	&2.21(0.17)
				\\ \hline
				\textbf{Error 2}  &TP	&14.48(2.04)	&23.06(1.96)	&14.42(2.12)	&15.18(2.06)	&24.02(1.48)	&15.48(2.30)
				\\
				&FP	&0.64(0.85)	&32.26(7.41)	&0.74(0.88)	&2.20(1.55)	&85.78(20.06)	&14.06(4.41)
				\\
				&Pred &2.57(0.11)	&2.85(0.13)	&2.57(0.11)	&12.43(1.15)	&15.92(1.68)	&16.55(1.69)
				\\ \hline
				\textbf{Error 3} &TP &13.74(2.65)	&22.52(2.38)	&13.80(2.66)	&14.30(2.70)	&23.92(1.37)	&14.62(2.69)
				\\
				&FP	&0.68(0.68)	&34.24(8.93)	&0.80(0.83)	&2.74(1.48)	&97.40(19.78)	&15.96(4.30)
				\\
				&Pred &2.71(0.12)	&3.00(0.14)	&2.71(0.12)	&14.36(1.35)	&18.52(1.70)	&19.25(1.84)
				\\ \hline
				\textbf{Error 4} &TP &16.90(3.12)	&21.83(3.04)	&16.90(3.36)	&11.70(5.86)	&20.70(5.74)	&12.07(5.44)	
				\\
				&FP &0.33(0.48)	&27.97(8.48)	&0.27(0.45)	&3.10(2.64)	&88.50(28.58)	&14.83(5.52)
				\\
				&Pred &1.85(0.15)	&2.10(0.17)	&1.85(0.15)	&16.25(9.88)	&22.78(17.05)	&24.20(18.91)
				\\  \hline
				\textbf{Error 5} &TP &16.26(2.28)	&23.42(2.01)	&16.42(2.16)	&13.80(3.37)	&23.24(2.25)	&14.24(3.05)
				\\
				&FP &0.32(0.62)	&29.38(7.54)	&0.32(0.65)	&3.00(2.14)	&94.72(27.12)	&16.26(4.84)
				\\
				&Pred &2.20(0.14)	&2.49(0.17)	&2.21(0.14)	&15.94(4.43)	&20.73(5.12)	&21.66(5.48)
				\\
				\hline
		\end{tabu} }
	\end{center}
	\centering
\end{table}

We have collected the posterior samples from the Gibbs sampler running 15,000 iterations while discarding the first 7,500 samples as burn-ins. The Bayesian estimates are calculated using the posterior medians. Simulation results for the gene expression data in Example 1 are tabulated in Table (\ref{perf.gene.ss}) and (\ref{perf.gene}). We can observe that the performance of methods that adopt spike--and--slab priors in Table (\ref{perf.gene.ss}) is consistently better than methods without spike--and--slab priors in Table (\ref{perf.gene}). Although, methods without spike--and--slab priors have slightly lower FPs than their counterparts with spike--and--slab priors under some error distributions, they tend to have much lower TPs and higher prediction errors under all the error distributions. For example, under Error2, RBSG identifies 14.48(SD 2.04) out of the 25 true positives, much lower than the true positives of 21.66(SD 1.72) from RBSG--SS. Meanwhile, its false positives 0.64(SD 0.85) is only slightly lower than the FP of RBSG--SS (1.32(SD 1.33)).  The prediction error of RBSG,  2.57 with a SD of 0.11, is also inferior than that of the RBSG--SS (2.15(SD 0.10)). Such an advantage can also be observed by comparing other methods in Table (\ref{perf.gene.ss}) with their counterparts (without spike--and--slab priors) from Table (\ref{perf.gene}). 

Among all the methods with spike--and--slab priors, as shown in Table (\ref{perf.gene.ss}), the proposed RBSG--SS has the best performance in both identification and prediction in the presence of data contamination and heavy--tailed errors. Under the mixture Laplace error (Error 3), RBSG--SS identifies 21.28(SD 2.24) true positives, with a small number of false positives, 1.48(SD 1.34). RBG--SS has a true positive of 24.80(SD 0.73), however, the number of false positives, 30.64(SD 4.23), is much higher. This is due to the fact that RBG--SS only conducts group level selection and does not impose the within-group sparsity. Compared to RBSG--SS, RBL--SS ignores the group structure, leading to fewer true positives of 18.14(SD2.68). In terms of prediction, RBSG--SS has the smallest L1 error, 2.29(0.12), among all the 3 robust methods with spike--and--slab priors. Although the difference in prediction error between RBSG--SS and RBG--SS is not distinct, considering the much smaller number of false positive main and interaction effects, we can fully observe the advantage of RBSG--SS over RBG--SS in prediction.       


Moreover, a cross--comparison between the robust and non--robust methods further demonstrates the necessity of developing robust Bayesian methods. For instance, under the error of $t$ distribution with 2 degrees of freedom (Error 4), RBSG--SS has identified 23.80(SD 1.30) true main and interaction effects with only 0.53(SD 0.86) false positives. Its direct non--robust competitor, BSG--SS, leads to a true positive of 16.20(SD 6.45) with 3.73(SD 4.61) false effects. The superior performance of RBSG--SS over the other two non--robust methods, BG--SS and BL--SS, is also clear. Although a comparison between the prediction errors of robust and non--robust methods is not feasible as the two are computed under the L1 and least square errors, the identification results convincingly suggest the advantage of robust methods over non-robust ones,    


Similar patterns have been observed in Table \ref{perf.qu.ss}, \ref{perf.qu}, \ref{perf.LD.ss}, \ref{perf.LD}, \ref{perf.T2D.ss} and \ref{perf.T2D} for Examples 2, 3 and 4, respectively, in the Appendix. Overall, based on the investigations over all the methods through comprehensive simulation studies, we can establish the advantage of conducting robust Bayesian bi--level selection incorporating spike--and--slab priors.    

We demonstrate the sensitivity of RBSG--SS for variable selection to the choice of the hyper--parameters for $\pi_{0}$, and $\pi_{1}$ in the Appendix. The results are tabulated in Table \ref{sens}, showing that the MPM model is insensitive to different specification of the hyper-parameters. {Following Li et al. \cite{LJH}, we assess the convergence of the MCMC chains by the potential scale reduction factor (PSRF) \cite{PSRF, PSRF2}. PSRF values close to 1 indicate that chains converge to the stationary distribution. Gelman et al. \cite{BDA} recommend using PSRF$\leq1.1$ as the cutoff for convergence, which has been adopted in our study. We compute the PSRF for each parameter and find the convergence of all chains after the burn-ins. For the purpose of demonstration, Figure \ref{fig:cvg} shows the pattern of PSRF the first five groups of coefficients in Example 1 under Error 2. The figure clearly shows the convergence of the proposed Gibbs sampler. }

\section{Real Data Analysis}
\label{makereference4.4}

\subsection{Nurses' Health Study (NHS) data}

Nurses' Health Study (NHS) is one of the largest investigations into the risk factors for major chronic diseases in women. As part of the Gene Environment Association Studies initiative (GENEVA), the NHS provides SNP genotypes data as well as detailed information on dietary and lifestyle variables. Obesity level is one of the most important risk factors for Type 2 diabetes mellitus (T2D), a chronic disease due to both genetic and environmental factors. In this study, we analyze the NHS type 2 diabetes data to identify main and interaction effects associated with obesity. We use weight as the response and focus on SNPs on chromosome 10. We consider five environment factors, including the total physical activity (act), glycemic load (gl), cereal fiber intake (ceraf), alcohol intake (alcohol) and a binary indicator of whether an individual has a history of high cholesterol (chol). All these environmental exposures have been suggested to be associated with obesity and diabetes \cite{T2D}. In addition, we include three clinical covariates: height, age and a binary indicator of whether an individual has a history of hypertension (hbp). In NHS study, about half of the subjects are diagnosed of type 2 diabetes and the other half are controls without the disease. We only use health subjects in this study. After cleaning the data through matching phenotypes and genotypes, removing SNPs with minor allele frequency (MAF) less than 0.05 or deviation from Hardy--Weinberg equilibrium, the working dataset contains 1732 subjects with 35099 SNPs.

For computational convenience prescreening can be conducted to reduce For computational convenience prescreening can be conducted to reduce the feature space to a more attainable size for variable selection. For example, Li et al. \cite{LJH} and Wu et al. \cite{WU2014} use the single SNP analysis to filter SNPs in a GWA study before downstream analysis. 
In this study, we use a marginal linear model with weight as the response variable to evaluate the penetrance effect of a variant under the environmental exposure. The marginal linear model uses a group of genetic main effect and G$\times$E interactions corresponding to a SNP as the predictors, and test whether this SNP has any effect, main or G$\times$E interaction. The SNPs with p-values less than a certain cutoff (0.001) for any effect, main or interaction, from the test are kept.
253 SNPs pass the screening.

The proposed approach RBSG-SS identifies 22 main SNP effects and 45 G$\times$E interactions. The detailed estimation results are provided in Table \ref{T2D.RBSG.SS} in the Appendix. We observe that the proposed method identifies main and interaction effects of SNPs with important implications in obesity. For example, two important SNPs, rs6482836 and rs10741150, that located within gene DOCK1 are identified. DOCK1 (Dedicator Of Cytokinesis 1) has been reported as a putative candidate for obesity related to adiponectin and triceps skinfold by previous studies \cite{DOCK1.1,DOCK1.2}. RBSG-SS identifies the main effect of rs6482836 and its interaction with the E factor act. Physical activity plays an important role in the prevention of overweight and obese \cite{act}. This result suggests that the expression level of DOCK1 in an individual may influence the effect of physical activity in obesity prevention.
RBSG-SS also identifies the interaction between rs10741150 and the E factor chol, suggesting that the effect of cholesterol level can be mediated by DOCK1. Interestingly, a previous study has shown that the expression level of DOCK5, an important paralog of DOCK1, is increased in individuals exposed to a diet high in saturated fatty acids \cite{DOCK1.3}. Our results provide more evidence of the importance of DOCK1 in diet-induced obesity. 
Another example is the SNP rs11196539, located within gene NRG3. NRG3(Neuregulin 3) has been found to be associated with both the basal metabolic rate (BMR) and body mass index (BMI) \cite{NRG3}. RBSG-SS identifies its interaction with the E factors, gl and alcohol. Both glycemic load and alcohol intake are important dietary variables associated with obesity. The continued intake of high-glycemic load meals leads to an increased risk of obesity \cite{GL}. The increasing alcohol consumption is associated with a decline in body mass index in women \cite{alcohol.1}, however, heavy drinking can increase risk of the metabolic syndrome \cite{alcohol.2}. Our results suggest that further investigation of NRG3 may help explain the mechanism of the effects of glycemic load and alcohol intake on obesity.
For the environment main effects, two E factors, chol and gl, have positive coefficients, and the other three, act, ceraf and alcohol, have negative coefficients, which are consistent with findings in the previous literature.

In addition to the proposed approach, we also conduct analysis using the alternatives RBL--SS, BSG--SS and BL--SS. As other alternative methods show inferior performance in simulation, they are not considered in real data analysis. Detailed estimation results are provided in Table \ref{T2D.RBL.SS}, \ref{T2D.BSG.SS} and \ref{T2D.BL.SS} in the Appendix. In Table \ref{simil}, we provide the numbers of identified main and interaction effects with pairwise overlaps, to show the difference in terms of identification between the proposed method and the others.
To further investigate the biological similarity of the identified genes, we conduct the Gene Ontology (GO) analysis. We can find an obvious difference between the proposed RBSG--SS and the three alternatives. The GO analysis results are provided in Figure \ref{fig:go}.

With real data, it is difficult to assess the selection accuracy objectively. The prediction performance may provide additional information to the selection results. Following Yan and Huang \cite{YJHJ} and Li et al. \cite{LJH},  we refit the models identified by RBSG--SS and RBL--SS using the robust Bayesian Lasso, and refit the models selected by BSG--SS and BL--SS using the Bayesian Lasso. For robust methods, the prediction mean absolute deviations (PMAD) are computed based on the posterior median estimates. The PMADs are 8.64 and 8.88 for RBSG--SS and RBL--SS, respectively. The proposed method outperforms the competitors. For non-robust methods, the prediction mean squared errors, or PMSEs, are 128.39 and 137.77 for BSG--SS and BL--SS, respectively. Overall, the superior performance of RNSG--SS over the alternatives can be observed. 

\subsection{TCGA skin cutaneous melanoma data}
In this case study, we analyze the Cancer Genome Atlas (TCGA) skin cutaneous melanoma (SKCM) data. TCGA is a collaborative effort supported by the National Cancer Institute (NCI) and the National Human Genome Research Institute (NHGRI), and has published high quality clinical, environmental, as well as multi--omics data. For this study, we use the level-3 gene expression data of SKCM downloaded from the cBio Cancer Genomics Portal \cite{cBio}. Our goal is to identify genes that have genetic main effect or G$\times$E interaction effects on the Breslow' thickness, an important prognostic variable for SKCM \cite{Breslow}. The log-transformed Breslow’s depth is used as the response variable and four E factors are considered, age, AJCC pathologic tumor stage, gender and Clark level. Data are available on 294 subjects and 20,531 gene expressions. We adopt the same screening method used in the first case study to select 109 genes  for further analysis.

The proposed approach RBSG--SS identifies 16 main SNP effects and 32 G$\times$E interactions. The detailed estimation results are available from Table \ref{SKCM.RBSG.SS} in the Appendix. One important gene identified is CXCL6 (C-X-C Motif Chemokine Ligand 6), a chemokine with neutrophil chemotactic and angiogenic activities. It has been reported that CXCL6 plays an important role in melanoma growth and metastasis \cite{CXCL6}. RBSG-SS identifies its main effect and its interactions with E factors, stage and Clark level. This suggests that CXCL6 can have different effects at different stages of melanoma. Another important finding is the gene MAGED4, one of member in MAGE(Melanoma-associated antigen) family. MAGE family contains genes that are highly attractive targets for cancer immunotherapy \cite{MAGED4}. MAGED4 has been found to be an potential target for glioma immunotherapy \cite{MAGE}. RBSG-SS identifies the main effect of MAGED4 and its interaction with the E factor tumor stage, suggesting that MAGED4 may also play an important role in SKCM and its effect may change over different tumor stages. 
For the main effects of the E factors, Clark level and tumor stage have positive coefficients, and age and gender have negative coefficients, which match observations in the literature.

Analysis is also conducted via the three alternative methods, and the results are summarized in Table \ref{simil}. Detailed estimation results are provided in Table \ref{SKCM.RBL.SS}, \ref{SKCM.BSG.SS} and \ref{SKCM.BL.SS} in the Appendix. Again, the proposed RBSG--SS identifies different sets of main and interaction effects from the rest. We further investigate the biological similarity of the identified genes by GO analysis (Figure \ref{fig:go}), which suggests an obvious difference. Prediction performance is also evaluated. The PMADs are 0.69 and 0.83 for RBSG--SS and RBL--SS, respectively. The proposed approach again has better prediction performance than RBL--SS. The PMSEs are 0.93 and 1.05 for BSG-SS and BL-SS, respectively. Combined, the RBSG--SS outperforms the alternatives.

\begin{table} [h!]
	\def\arraystretch{1.5}
	\begin{center}
		\caption[Overlaps in NHS T2D.]{The numbers of main G effects and interactions identified by different approaches and their overlaps.}\label{simil}
		\centering
		\fontsize{10}{12}\selectfont{
			\begin{tabu} to \textwidth{ X[1.1l] X[1.1c] X[c] X[c] X[c] X[1.1c] X[c] X[c] X[c]}
				\hline
				\textbf{NHS} &  \multicolumn{4}{c}{Main G effects}  & \multicolumn{4}{c}{Interactions}  \\
				\cmidrule(lr){2-5} \cmidrule(lr){6-9}
				&RBSG-SS &  RBL-SS  & BSG-SS & BL-SS  &RBSG-SS &  RBL-SS  & BSG-SS & BL-SS   \\
				\hline
				{RBSG-SS} &22 &20 &16 &13 &45 &21 &17 &10			 		
				\\
				{RBL-SS}  & &29 &20 &16 & &39 &14 &14
				\\
				{BSG-SS} & & &29 &25 & & &34 &22
				\\
				{BL-SS} &&&&27 &&&&42
				\\
				\hline
				\textbf{SKCM} &  \multicolumn{4}{c}{Main G effects}  & \multicolumn{4}{c}{Interactions}  \\
				\cmidrule(lr){2-5} \cmidrule(lr){6-9}
				&RBSG-SS &  RBL-SS  & BSG-SS & BL-SS  &RBSG-SS &  RBL-SS  & BSG-SS & BL-SS   \\
				\hline
				{RBSG-SS} &16 &10 &14 &13 &32 &11 &18 &10			 		
				\\
				{RBL-SS}  & &17 &12 &14 & &33 &15 &24
				\\
				{BSG-SS} & & &22 &15 & & &29 &14
				\\
				{BL-SS} &&&&20 &&&&33
				\\
				\hline
		\end{tabu} }
	\end{center}
	\centering
\end{table}

\section{Discussion}
\label{makereference4.5}
In this study, we have developed robust Bayesian variable selection methods for gene-environment interaction studies. The robustness of our methods comes from Bayesian formulation of LAD regression. In G$\times$E studies, the demand for robustness arises in heavy-tailed distribution/ data contamination in both the response and predictors, as well as model misspecification. We have focused on the first case, which is frequently encountered in practice. Investigations of the robust Bayesian methods accommodating the other two cases are interesting and will be pursued in the future.

Recently, penalization has emerged as a power tool for dissecting G$\times$E interactions \cite{ZHOU}. Our literature review suggests that Bayesian variable selection methods, although tightly related to penalization, has not been fully explored for interaction analyses, let alone the robust ones.  We are among the first to conduct robust G$\times$E analysis within the Bayesian framework. The proposed Bayesian LAD sparse group LASSO are not only specifically tailored for G$\times$E studies, and but also generally applicable for problems incorporating the bi-level structure in a broader context, such as simultaneously selection of prognostic genes and pathways \cite{JYU,LWOS}. The spike-and-slab priors have been incorporated to further improve identification and prediction performances. As a byproduct, the Bayesian LAD LASSO and group LASSO, both with spike-and-slab priors, have also been investigated for the first time. The computational feasibility of the Gibbs samplers is guaranteed by the R package roben, with the core modules of the MCMC algorithms developed in C++. 

In G$\times$E studies, the form of interaction effects can be linear, nonlinear, and both linear and nonlinear, resulting in parametric \cite{WU2018SM,XYQ,LIPID}, nonparametric \cite{WU2013,LJH,WU2018} and semiparametric variable selection methods \cite{WU2014, WU2015, REN2020} to dissect G$\times$E interactions, respectively. The proposed study can be potentially generalized to these studies within robust Bayesian framework. For example, variable selection for multiple semiparametric G$\times$E studies can be formulated as a combination of individual and group level selection problem, where the robust Bayesian methods based on sparse group, group and individual level selection are directly applicable. The proposed robust Bayesian framework has paved the way for the future investigations.



\bibliography{references}

\clearpage
\appendix

\section{Summary of methods.}

\begin{table} [h!]
	\def\arraystretch{1.5}
	\begin{center}
		\caption[Summary of the proposed and alternative methods.]{Summary of the proposed and alternative methods.}\label{methods}
		\centering
		\fontsize{10}{12}\selectfont{
			\begin{tabu} to \textwidth{ X[0.3l] X[0.3l] X[l] X[0.6l]}
				\hline
				&     & \textbf{Methods}  & \textbf{Reference}     \\
				\hline
				\multirow{6}{5em}{\textbf{Robust}} &\textbf{RBSG-SS} & Robust Bayesian sparse group selection with spike--and--slab priors & proposed for the first time
				\\
				&\textbf{RBG-SS} & Robust Bayesian group selection with spike--and--slab priors & proposed for the first time
				\\
				&\textbf{RBL-SS} &  Robust Bayesian Lasso with spike--and--slab priors & proposed for the first time
				\\
				\cmidrule(lr){2-4}
				&\textbf{RBSG} & Robust Bayesian sparse group selection & proposed for the first time
				\\
				&\textbf{RBG} & Robust Bayesian group Lasso & Li et al. (2010) \cite{BRQR}
				\\
				&\textbf{RBL} & Robust Bayesian Lasso & Li et al. (2010) \cite{BRQR}
				\\
				\hline
				\multirow{6}{10em}{\textbf{Non-robust}} &\textbf{BSG-SS} & Bayesian sparse group Lasso with spike--and--slab priors & Xu and Ghosh (2015) \cite{XXF}
				\\
				&\textbf{BG-SS} & Bayesian group Lasso with spike--and--slab priors & Xu and Ghosh (2015)\cite{XXF} Zhang et al. (2014) \cite{ZLIN}
				\\
				&\textbf{BL-SS} &  Bayesian Lasso with spike--and--slab priors & Xu and Ghosh (2015)\cite{XXF} Zhang et al. (2014)\cite{ZLIN}
				\\
				\cmidrule(lr){2-4}
				&\textbf{BSG} & Bayesian sparse group Lasso & Xu and Ghosh (2015) \cite{XXF}
				\\
				&\textbf{BG} & Bayesian group Lasso & Kyung et al. (2010) \cite{KYU}
				\\
				&\textbf{BL} & Bayesian Lasso & Park and Casella (2008) \cite{PARK}
				\\
				\hline
				\multicolumn{4}{l}{\footnotesize Note: The models in the references are modified to be applicable to G$\times$E settings.} \\
		\end{tabu} }
	\end{center}
\end{table}

\clearpage


\section{Hyper-parameters sensitivity analysis}
We demonstrate the sensitivity of RBSG-SS for variable selection to the choice of the hyperparameters for $\pi_{0}$, and $\pi_{1}$. We consider five different Beta priors: (1) Beta(0.5, 0.5) which is a U-shape curve between $(0,1)$; (2) Beta(1, 1) which is a essentially a uniform prior; (3) Beta(2, 2) which is a quadratic curve; (4) Beta(1, 5) which is highly right-skewed; (5) Beta(5, 1) which is highly left-skewed. As a demonstrating example, we use the same setting of Example 1 to generate data under the Error 2. Table \ref{sens} shows the identification performance of the median thresholding model (MPM) with different Beta priors. For all choices of Beta priors, the MPM model is very stable. Also, RBSG-SS correctly identifies most of the true effects with low false positives in all cases. {Therefore, we simply use Beta$(1,1)$ as the prior for $\pi_{0}$, and $\pi_{1}$ in this study.}

\begin{table} [h!]
	\def\arraystretch{1.5}
	\begin{center}
		\caption[Sensitivity analysis on hyper-parameters.]{Sensitivity analysis for RBSG-SS using Example 1. mean(sd) of true positives (TP), false positives (FP) and prediction errors (Pred) based on 100 replicates.}\label{sens}
		\centering
		\fontsize{10}{12}\selectfont{
			\begin{tabu} to 0.6\textwidth{ X[1.3l] X[c] X[c] X[c] }
				\hline
				&  TP  & FP & Pred
				\\
				\hline
				\textbf{Beta}(0.5, 0.5) &21.31(1.67)	&1.71(1.50)	&2.19(0.11)
				\\
				\textbf{Beta}(1, 1)  &21.66(1.72)	&1.32(1.33)	&2.17(0.10)
				\\
				
				\textbf{Beta}(2, 2) &21.13(2.10)	&1.47(1.16)	&2.18(0.10)
				\\
				\textbf{Beta}(1, 5) &20.82(1.71)	&1.38(1.30)	&2.17(0.10)
				\\
				\textbf{Beta}(5, 1) &21.58(1.75)	&2.22(1.52)	&2.19(0.09)
				\\
				\hline
		\end{tabu} }
	\end{center}
	\centering
\end{table}

\clearpage
\section{{Assessment of the convergence of MCMC chains}}
\begin{figure}[h!]
	\centering
	\includegraphics[angle=0,origin=c,width=1\textwidth]{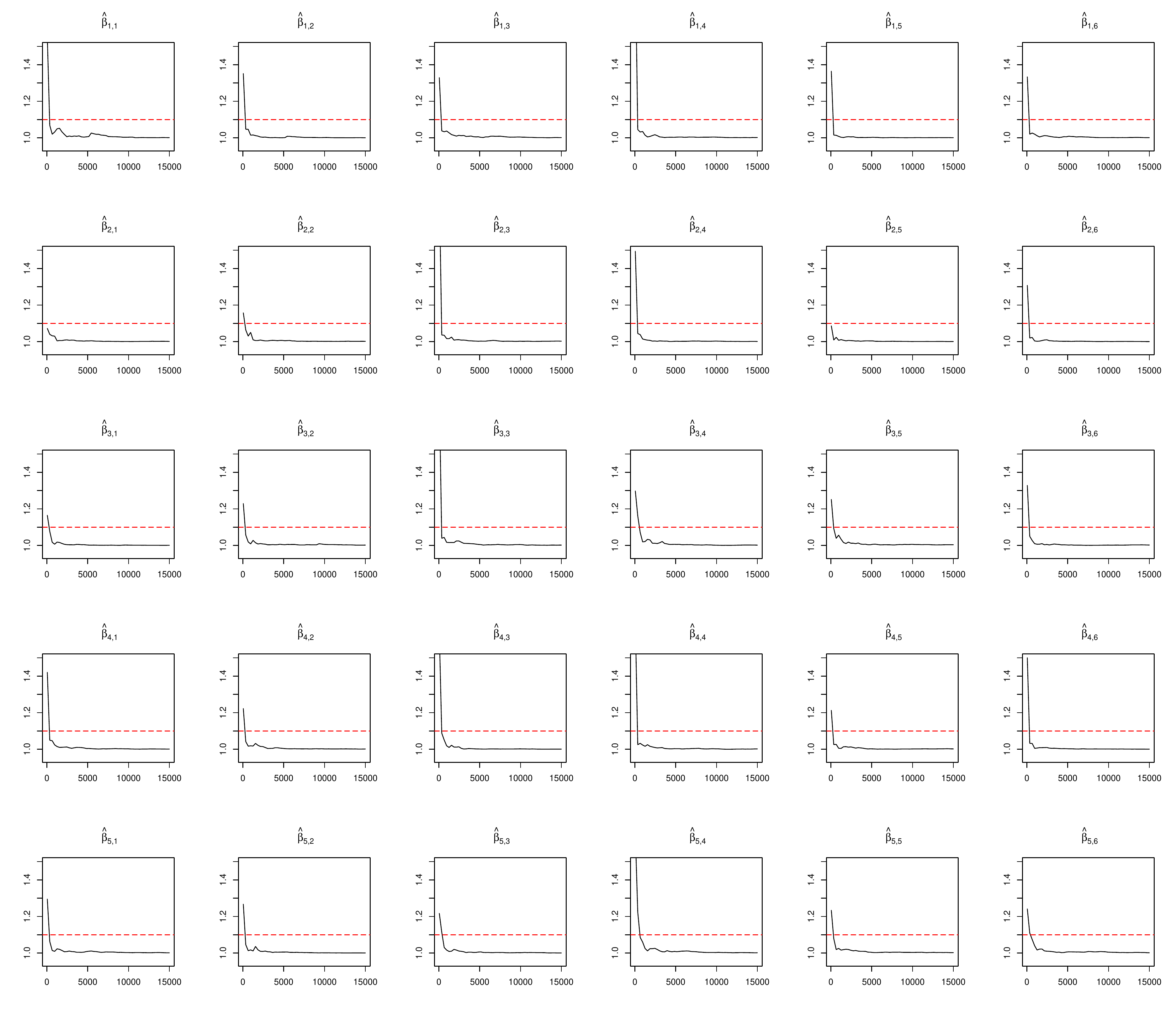}
	\caption[Potential scale reduction factor against iterations]{{Potential scale reduction factor (PSRF) against iterations for the first five groups of coefficients in Example 1. Black line: the PSRF. 
			Red line: the threshold of 1.1. The $\hat{\beta}_{j1}$ to $\hat{\beta}_{j6}$ represent the six estimated coefficients for the main and interaction effects in the $j$th group, $(j=0,\dots,5)$, respectively.}}
	\label{fig:cvg}
\end{figure}

\clearpage

\section{Additional simulation results}

\begin{table} [h!]
	\def\arraystretch{1.5}
	\begin{center}
		\caption[Simulation results in Example 2.]{Simulation results in Example 2. $(n,q,k,p)$ = (500, 2, 5, 100). mean(sd) of true positives (TP), false positives (FP)and prediction errors (Pred) based on 100 replicates.}\label{perf.qu.ss}
		\centering
		\fontsize{10}{12}\selectfont{
			\begin{tabu} to \textwidth{ X[0.8c] X[0.5r] X[c] X[c] X[c] | X[c] X[c] X[c]}
				\hline
				&&  RBSG-SS  & RBG-SS & RBL-SS  & BSG-SS    & BG-SS & BL-SS      \\
				\cmidrule(lr){1-8}
				\textbf{Error 1} &TP &24.87(0.35)	&25.00(0.00)	&24.53(0.51)	&24.83(0.38)	&25.00(0.00)	&24.53(0.51)
				\\
				{}&FP	&1.63(1.16)	&31.40(3.38)	&2.30(1.86)	&1.13(1.04)	&29.20(1.10)	&0.60(0.85)
				\\
				&Pred	&0.85(0.03)	&0.86(0.03)	&0.86(0.03)	&1.09(0.06)	&1.13(0.06)	&1.10(0.07)
				\\ \hline
				\textbf{Error 2}  &TP &22.23(1.76)	&24.67(0.76)	&19.23(1.72)	&19.97(1.63)	&24.47(0.90)	&15.27(1.91)
				\\
				{}&FP	&1.90(1.30)	&35.73(7.83)	&2.10(1.40)	&2.33(1.42)	&34.13(7.44)	&1.73(1.39)
				\\
				&Pred &2.24(0.14)	&2.18(0.11)	&2.38(0.16)	&10.21(1.27)	&9.13(0.94)	&11.30(1.83)
				\\ \hline
				\textbf{Error 3} &TP &21.50(1.48)	&25.00(0.00)	&17.43(2.13)	&18.73(2.02)	&25.00(0.00)	&13.10(1.54)				
				\\
				{}&FP &2.13(1.14)	&35.20(6.77)	&1.90(1.37)	&2.90(1.71)	&34.00(6.88)	&1.37(0.96)
				\\
				&Pred &2.39(0.18)	&2.29(0.11)	&2.52(0.22)	&12.46(1.67)	&10.40(0.94)	&13.04(1.35)
				\\ \hline
				\textbf{Error 4} &TP &23.58(1.49)	&25.00(0.00)	&21.04(2.29)	&15.94(5.34)	&23.18(3.50)	&12.08(4.60)
				\\
				{}&FP &0.80(0.93)	&30.32(3.27)	&0.78(1.07)	&7.46(27.02)	&53.50(58.05)	&3.56(8.91)
				\\
				&Pred &1.85(0.16)	&1.82(0.13)	&1.92(0.17)	&25.65(55.13)	&25.63(67.60)	&30.67(87.77)
				\\ \hline
				\textbf{Error 5} &TP &24.12(1.00)	&25.00(0.00)	&21.82(1.90)	&18.04(3.64)	&24.24(1.88)	&13.12(2.99)
				\\
				{}&FP &0.90(1.02)	&29.48(1.64)	&0.82(0.90)	&2.72(1.75)	&36.12(12.21)	&1.48(1.25)
				\\
				&Pred &1.81(0.13)	&1.82(0.12)	&1.89(0.15)	&14.85(6.53)	&12.87(5.94)	&15.19(6.43)
				\\
				\hline
		\end{tabu} }
	\end{center}
	\centering
\end{table}

\begin{table} [h!]
	\def\arraystretch{1.5}
	\begin{center}
		\caption[Simulation results in Example 2 (2).]{Simulation results in Example 2. $(n,q,k,p)$ = (500, 2, 5, 100). mean(sd) of true positives (TP), false positives (FP) and prediction errors (Pred) based on 100 replicates.}\label{perf.qu}
		\centering
		\fontsize{10}{12}\selectfont{
			\begin{tabu} to \textwidth{ X[0.8c] X[0.5r] X[c] X[c] X[c] | X[c] X[c] X[c]}
				\hline
				&&  RBSG  & RBG & RBL  & BSG    & BG & BL     \\
				\cmidrule(lr){1-8}
				\textbf{Error 1} &TP &24.20(0.61)	&25.00(0.00)	&24.23(0.57)	&24.33(0.61)	&25.00(0.00)	&24.30(0.60)
				\\
				{}&FP	&2.93(1.86)	&54.80(17.34)	&3.30(1.97)	&1.87(1.61)	&56.20(10.30)	&5.77(2.56)
				\\
				&Pred	&1.14(0.05)	&1.32(0.06)	&1.15(0.05)	&1.74(0.12)	&2.13(0.14)	&2.01(0.15)
				\\ \hline
				\textbf{Error 2}  &TP &14.00(2.27)	&22.20(2.33)	&13.63(2.66)	&13.70(2.29)	&23.63(1.61)	&14.20(1.97)
				\\
				{}&FP	&0.60(0.85)	&31.40(12.07)	&0.83(1.05)	&1.33(1.18)	&62.77(24.90)	&8.80(4.54)
				\\
				&Pred &2.57(0.13)	&2.77(0.14)	&2.58(0.14)	&12.18(1.15)	&14.42(1.40)	&14.91(1.43)
				\\ \hline
				\textbf{Error 3} &TP &12.40(2.03)	&22.47(1.17)	&12.27(1.87)	&12.43(1.77)	&23.20(1.49)	&13.37(2.13)			
				\\
				{}&FP &0.57(0.77)	&29.33(5.54)	&0.60(0.93)	&1.47(1.31)	&59.80(13.17)	&8.17(3.04)
				\\
				&Pred &2.69(0.11)	&2.86(0.11)	&2.69(0.10)	&13.59(1.05)	&16.32(1.46)	&16.93(1.60)
				\\ \hline
				\textbf{Error 4} &TP &15.98(2.92)	&23.04(2.78)	&16.10(3.12)	&10.20(5.31)	&20.52(5.81)	&11.08(5.00)
				\\
				{}&FP &0.26(0.53)	&27.36(6.21)	&0.30(0.65)	&2.34(3.56)	&65.04(30.70)	&9.38(6.26)
				\\
				&Pred &2.19(0.16)	&2.35(0.16)	&2.21(0.17)	&26.27(53.26)	&34.08(78.47)	&34.95(79.04)
				\\ \hline
				\textbf{Error 5} &TP &16.48(2.69)	&23.48(1.58)	&16.30(2.63)	&11.96(3.66)	&22.26(3.70)	&12.70(3.50)
				\\
				{}&FP &0.32(0.59)	&28.72(5.89)	&0.34(0.59)	&1.40(1.21)	&62.34(20.52)	&8.34(3.77)
				\\
				&Pred &2.20(0.14)	&2.41(0.14)	&2.20(0.13)	&15.79(5.97)	&18.90(6.61)	&19.53(6.65)
				\\
				\hline
		\end{tabu} }
	\end{center}
	\centering
\end{table}

\begin{table} [h!]
	\def\arraystretch{1.5}
	\begin{center}
		\caption[Simulation results in Example 3.]{Simulation results in Example 3. $(n,q,k,p)$ = (500, 2, 5, 100). mean(sd) of true positives (TP), false positives (FP)and prediction errors (Pred) based on 100 replicates.}\label{perf.LD.ss}
		\centering
		\fontsize{10}{12}\selectfont{
			\begin{tabu} to \textwidth{ X[0.8c] X[0.5r] X[c] X[c] X[c] | X[c] X[c] X[c]}
				\hline
				&&  RBSG-SS  & RBG-SS & RBL-SS  & BSG-SS    & BG-SS & BL-SS      \\
				\cmidrule(lr){1-8}
				\textbf{Error 1} &TP &24.00(0.91)	&25.00(0.00)	&22.13(1.57)	&24.33(0.66)	&25.00(0.00)	&22.83(1.37)
				\\
				{}&FP	&1.85(1.46)	&33.40(5.21)	&1.87(1.36)	&1.27(1.34)	&29.60(1.83)	&0.77(1.04)
				\\
				&Pred	&0.86(0.03)	&0.86(0.03)	&0.89(0.03)	&1.11(0.07)	&1.13(0.07)	&1.17(0.10)
				\\ \hline
				\textbf{Error 2}  &TP &17.63(2.37)	&24.73(0.69)	&14.37(2.54)	&15.00(2.32)	&23.73(1.46)	&11.00(1.95)
				\\
				{}&FP	&2.50(1.41)	&33.27(5.14)	&2.67(1.99)	&2.60(1.43)	&30.67(6.69)	&1.87(1.53)
				\\
				&Pred &2.33(0.12)	&2.15(0.10)	&2.40(0.17)	&10.37(1.01)	&9.01(0.81)	&10.71(0.94)
				\\ \hline
				\textbf{Error 3} &TP &17.23(1.77)	&24.80(0.61)	&14.47(2.21)	&15.10(2.29)	&23.67(1.77)	&11.03(1.38)
				\\
				{}&FP	&2.27(1.78)	&32.20(6.94)	&1.63(1.43)	&2.13(1.70)	&30.93(5.48)	&1.17(1.34)
				\\
				&Pred &2.39(0.13)	&2.24(0.10)	&2.45(0.13)	&11.98(1.45)	&10.32(1.04)	&12.37(1.41)	
				\\ \hline
				\textbf{Error 4} &TP &23.63(1.19)	&24.67(0.92)	&20.13(2.19)	&15.07(4.69)	&22.67(3.68)	&11.40(4.01)
				\\
				{}&FP &1.30(1.12)	&29.13(2.67)	&1.17(0.95)	&3.37(1.88)	&29.93(9.48)	&2.37(1.97)
				\\
				&Pred &1.48(0.13)	&1.45(0.11)	&1.55(0.14)	&12.66(12.40)	&10.10(8.77)	&12.75(11.83)
				\\ \hline
				\textbf{Error 5} &TP &24.80(0.48)	&25.00(0.00)	&23.57(1.43)	&20.30(2.83)	&24.87(0.51)	&15.87(2.32)
				\\
				{}&FP &0.33(0.55)	&29.60(1.83)	&0.40(1.04)	&3.00(1.66)	&32.93(5.86)	&2.33(1.63)
				\\
				&Pred &1.19(0.10)	&1.21(0.10)	&1.21(0.11)	&6.055(1.77)	&5.47(1.73)	&6.54(1.75)
				\\
				\hline
		\end{tabu} }
	\end{center}
	\centering
\end{table}

\begin{table} [h!]
	\def\arraystretch{1.5}
	\begin{center}
		\caption[Simulation results in Example 3 (2).]{Simulation results in Example 3. $(n,q,k,p)$ = (500, 2, 5, 100). mean(sd) of true positives (TP), false positives (FP) and prediction errors (Pred) based on 100 replicates.}\label{perf.LD}
		\centering
		\fontsize{10}{12}\selectfont{
			\begin{tabu} to \textwidth{ X[0.8c] X[0.5r] X[c] X[c] X[c] | X[c] X[c] X[c]}
				\hline
				&&  RBSG  & RBG & RBL  & BSG    & BG & BL      \\
				\cmidrule(lr){1-8}
				\textbf{Error 1} &TP &21.27(1.17)	&25.00(0.00)	&21.30(1.06)	&22.23(0.94)	&25.00(0.00)	&22.13(1.28)
				\\
				{}&FP	&1.97(1.56)	&45.40(11.68)	&2.03(1.56)	&1.23(1.33)	&43.60(10.41)	&3.37(2.03)
				\\
				&Pred	&1.08(0.04)	&1.21(0.05)	&1.07(0.04)	&1.57(0.12)	&1.87(0.13)	&1.79(0.13)
				\\ \hline
				\textbf{Error 2}  &TP &8.40(1.87)	&18.67(2.68)	&8.27(2.02)	&8.07(1.57)	&20.73(2.65)	&8.73(2.15)
				\\
				{}&FP	&0.43(0.63)	&20.73(4.93)	&0.57(0.77)	&0.67(0.66)	&36.27(11.59)	&3.83(2.05)
				\\
				&Pred &2.44(0.13)	&2.58(0.13)	&2.44(0.12)	&10.97(1.05)	&12.78(1.31)	&13.02(1.33)
				\\ \hline
				\textbf{Error 3} &TP &8.43(2.18)	&16.70(3.29)	&8.70(2.00)	&7.97(2.04)	&18.60(3.07)	&8.27(1.78)
				\\
				{}&FP	&0.33(0.71)	&17.70(4.60)	&0.43(0.73)	&0.60(0.72)	&33.60(10.63)	&3.70(2.34)
				\\
				&Pred &2.54(0.11)	&2.69(0.12)	&2.55(0.11)	&12.33(1.15)	&14.30(1.40)	&14.55(1.40)	
				\\ \hline
				\textbf{Error 4} &TP &13.77(2.18)	&21.20(2.06)	&13.67(2.04)	&9.67(3.74)	&20.60(4.79)	&9.77(3.76)
				\\
				{}&FP &0.43(0.63)	&22.80(3.98)	&0.57(0.63)	&1.03(1.13)	&38.00(12.21)	&4.47(2.50)
				\\
				&Pred &1.73(0.12)	&1.85(0.13)	&1.73(0.13)	&11.78(9.05)	&13.94(11.84)	&14.22(12.41)
				\\ \hline
				\textbf{Error 5} &TP &19.10(1.86)	&24.87(0.73)	&19.10(1.60)	&15.27(2.94)	&24.07(1.70)	&15.07(2.88)
				\\
				{}&FP &0.20(0.48)	&31.13(5.96)	&0.23(0.57)	&1.10(1.16)	&43.93(11.82)	&3.83(2.21)
				\\
				&Pred &1.45(0.08)	&1.61(0.09)	&1.46(0.08)	&6.13(1.13)	&7.19(1.24)	&7.16(1.29)
				\\
				\hline
		\end{tabu} }
	\end{center}
	\centering
\end{table}

\begin{table} [h!]
	\def\arraystretch{1.5}
	\begin{center}
		\caption[Simulation results in Example 4.]{Simulation results in Example 4. $(n,q,k,p)$ = (500, 2, 5, 100). mean(sd) of true positives (TP), false positives (FP) and prediction errors (Pred) based on 100 replicates.}\label{perf.T2D.ss}
		\centering
		\fontsize{10}{12}\selectfont{
			\begin{tabu} to \textwidth{ X[0.8c] X[0.5r] X[c] X[c] X[c] | X[c] X[c] X[c]}
				\hline
				&&  RBSG-SS  & RBG-SS & RBL-SS  & BSG-SS    & BG-SS & BL-SS      \\
				\cmidrule(lr){1-8}
				\textbf{Error 1} &TP &24.93(0.37)	&25.00(0.00)	&24.93(0.25)	&25.00(0.00)	&25.00(0.00)	&24.90(0.31)
				\\
				{}&FP	&1.33(0.99)	&30.60(3.84)	&1.47(1.25)	&1.00(1.02)	&29.20(1.10)	&0.33(0.61)
				\\
				&Pred	&0.84(0.02)	&0.88(0.02)	&0.85(0.03)	&1.10(0.04)	&1.20(0.06)	&1.11(0.05)
				\\ \hline
				\textbf{Error 2}  &TP &20.80(2.65)	&23.60(1.47)	&17.24(2.96)	&18.58(3.46)	&23.04(1.64)	&14.08(3.26)
				\\
				{}&FP	&1.32(1.22)	&30.76(4.93)	&1.66(1.29)	&1.98(1.53)	&27.72(5.49)	&1.42(1.25)
				\\
				&Pred &2.25(0.11)	&2.22(0.08)	&2.37(0.12)	&10.32(1.25)	&9.53(0.75)	&11.43(1.20)
				\\ \hline
				\textbf{Error 3} &TP &20.56(2.73)	&23.69(1.38)	&16.53(3.20)	&17.56(3.49)	&22.80(1.65)	&12.67(3.39)
				\\
				{}&FP &1.40(1.30)	&30.04(5.46)	&1.78(1.82)	&1.76(1.28)	&27.60(5.24)	&1.22(1.43)
				\\
				&Pred &2.38(0.13)	&2.35(0.10)	&2.51(0.16)	&12.04(1.40)	&11.12(0.96)	&13.32(1.44)
				\\ \hline
				\textbf{Error 4} &TP &24.60(0.93)	&24.67(0.92)	&23.77(1.57)	&20.10(6.38)	&22.27(5.10)	&15.63(6.69)
				\\
				{}&FP &0.40(0.56)	&29.13(2.97)	&0.47(0.73)	&1.83(1.90)	&28.13(9.22)	&1.17(1.15)
				\\
				&Pred &1.48(0.09)	&1.52(0.09)	&1.51(0.11)	&11.54(6.94)	&11.33(6.95)	&12.64(6.74)
				\\ \hline
				\textbf{Error 5} &TP &23.16(1.68)	&24.96(0.28)	&19.60(2.14)	&15.64(3.76)	&23.44(1.83)	&11.60(2.75)
				\\
				{}&FP &1.08(1.16)	&29.16(1.33)	&0.72(0.83)	&2.20(1.83)	&30.56(7.43)	&1.48(1.43)
				\\
				&Pred &1.56(0.14)	&1.53(0.13)	&1.63(0.15)	&10.98(5.80)	&9.45(5.38)	&11.38(6.03)
				\\
				\hline
		\end{tabu} }
	\end{center}
	\centering
\end{table}

\begin{table} [h!]
	\def\arraystretch{1.5}
	\begin{center}
		\caption[Simulation results in Example 4 (2).]{Simulation results in Example 4. $(n,q,k,p)$ = (500, 2, 5, 100). mean(sd) of true positives (TP), false positives (FP) and prediction errors (Pred) based on 100 replicates.}\label{perf.T2D}
		\centering
		\fontsize{10}{12}\selectfont{
			\begin{tabu} to \textwidth{ X[0.8c] X[0.5r] X[c] X[c] X[c] | X[c] X[c] X[c]}
				\hline
				&&  RBSG  & RBG & RBL  & BSG    & BG & BL      \\
				\cmidrule(lr){1-8}
				\textbf{Error 1} &TP &21.47(1.87)	&24.40(1.07)	&21.67(1.81)	&22.70(1.64)	&24.87(0.51)	&22.53(1.85)
				\\
				{}&FP	&3.17(2.51)	&56.00(20.03)	&3.33(2.59)	&2.30(1.66)	&66.33(14.04)	&6.57(2.62)
				\\
				&Pred	&1.26(0.06)	&1.44(0.07)	&1.27(0.06)	&2.40(0.27)	&2.83(0.24)	&2.75(0.30)
				\\ \hline
				\textbf{Error 2}  &TP &9.08(2.54)	&19.38(3.12)	&9.20(2.60)	&9.68(2.41)	&20.82(2.93)	&10.80(2.65)
				\\
				{}&FP	&0.78(0.86)	&30.30(10.26)	&0.84(0.89)	&2.18(1.48)	&65.94(19.60)	&8.62(3.38)
				\\
				&Pred &2.67(0.08)	&2.89(0.09)	&2.67(0.09)	&13.54(0.87)	&16.38(1.19)	&16.87(1.27)
				\\ \hline
				\textbf{Error 3} &TP &8.51(2.31)	&18.71(3.37)	&8.62(2.33)	&9.02(2.33)	&20.60(2.76)	&10.58(2.50)
				\\
				{}&FP &0.56(0.69)	&25.29(7.94)	&0.56(0.66)	&1.87(1.36)	&56.87(15.55)	&7.38(2.91)
				\\
				&Pred &2.79(0.11)	&3.00(0.12)	&2.79(0.12)	&15.34(1.29)	&18.66(1.69)	&19.30(1.88)
				\\ \hline
				\textbf{Error 4} &TP &13.30(3.32)	&21.93(2.72)	&13.47(3.33)	&10.93(4.30)	&20.97(4.76)	&11.97(4.43)
				\\
				{}&FP &0.50(0.57)	&29.07(9.28)	&0.40(0.50)	&1.70(1.39)	&60.03(20.56)	&7.90(3.92)
				\\
				&Pred &2.03(0.12)	&2.22(0.12)	&2.03(0.12)	&15.20(7.98)	&18.61(12.19)	&19.41(13.39)
				\\ \hline
				\textbf{Error 5} &TP &14.38(2.64)	&22.36(2.22)	&14.40(2.70)	&10.12(3.53)	&20.84(3.74)	&10.56(3.39)
				\\
				{}&FP &0.30(0.58)	&25.16(4.36)	&0.22(0.46)	&0.88(1.12)	&36.52(12.60)	&3.70(2.57)
				\\
				&Pred &1.84(0.15)	&1.99(0.17)	&1.84(0.15)	&11.23(5.54)	&13.02(5.80)	&13.18(5.90)
				\\
				\hline
		\end{tabu} }
	\end{center}
	\centering
\end{table}

\clearpage
\section{Estimation results for data analysis}

\begin{longtable}{ l l c c c  c c c}
	\caption[Analysis of the NHS T2D data using RBSG-SS.]{Analysis of the NHS T2D data using RBSG-SS.} \label{T2D.RBSG.SS} \\
	\hline
	SNP & $\text{Gene}^\ast$ &    & chol & act  & gl    & ceraf & alcohol \\ \hline 
	\endfirsthead
	\caption[]{Continued from the previous page.} \\ \hline
	SNP & $\text{Gene}^\ast$ &    & chol & act  & gl    & ceraf & alcohol \\ \hline 
	\endhead
	\hline 
	\multicolumn{8}{r@{}}{Continued on the next page}\\
	\endfoot
	\hline
	\multicolumn{8}{l@{}}{$\ast$ Genes that SNPs belong to or are the closest to.}\\
	\endlastfoot
	& &  &3.503 &-3.447  &0.752  & -3.364 &-2.639
	\\
	rs10741150	&DOCK1	&	&-0.948	&	&	&	&
	\\
	rs10765059	&TCERG1L	&-0.531	&	&	&	&	&0.877
	\\
	rs10786611	&RF00019	&0.668	&0.723	&	&0.530	&	&
	\\
	rs10884466	&RNA5SP326	&-0.466	&	&0.643	&	&	&
	\\
	rs10885423	&NRG3	&	&	&	&-0.715	&	&
	\\
	rs10886442	&GRK5	&	&0.805	&	&	&	&
	\\
	rs11196539	&NRG3	&	&	&	&-0.608	&	&-0.801
	\\
	rs11198590	&CACUL1	&-0.494	&	&	&	&0.994	&-0.687
	\\
	rs11259039	&FRMD4A	&1.016	&	&	&	&	&
	\\
	rs1194657	&THAP12P3	&	&	&	&0.798	&	&
	\\
	rs1219508	&RPS15AP5	&-0.742	&	&	&	&	&
	\\
	rs12265854	&SLC16A12	&0.397	&	&	&	&	&
	\\
	rs12414552	&TCERG1L	&0.667	&	&0.470	&0.585	&	&-0.690
	\\
	rs12767723	&SLC25A18P1	&0.820	&	&	&	&-0.515	&
	\\
	rs12772559	&TACR2	&	&	&	&0.938	&0.510	&
	\\
	rs12774333	&LRMDA	&-0.599	&	&	&	&	&
	\\
	rs12775160	&FOXI2	&-0.651	&-0.501	&	&	&	&0.647
	\\
	rs16916794	&SLC39A12	&-0.552	&0.511	&0.455	&	&	&
	\\
	rs16920092	&PLXDC2	&	&	&	&	&	&-0.843
	\\
	rs17094114	&GFRA1	&-0.615	&	&	&	&	&
	\\
	rs2492664	&OR6L1P	&0.695	&	&	&-0.737	&	&
	\\
	rs2784767	&PLAC9	&-0.540	&	&	&	&	&
	\\
	rs2814322	&GRID1	&	&	&	&-0.830	&	&
	\\
	rs3740063	&ABCC2	&	&	&	&-0.966	&	&
	\\
	rs3763722	&LARP4B	&0.332	&-1.156	&	&	&	&0.866
	\\
	rs4411238	&PRKG1	&0.537	&	&	&	&	&
	\\
	rs4578341	&CHST15	&	&-0.822	&	&	&0.602	&
	\\
	rs4747517	&ITIH5	&	&-1.468	&	&	&0.920	&
	\\
	rs4749926	&IL2RA	&-0.840	&-0.815	&	&	&	&
	\\
	rs4917817	&PYROXD2	&-0.624	&	&0.594	&	&	&
	\\
	rs4918904	&XRCC6P1	&	&	&	&	&0.997	&
	\\
	rs6482836	&DOCK1	&-0.957	&	&1.067	&	&	&
	\\
	rs7070789	&GPAM	&	&	&	&-1.245	&	&-0.791
	\\
	rs7072255	&ANTXRLP1	&	&0.800	&	&	&	&
	\\
	rs7077721	&SNRPD2P1	&0.858	&	&0.774	&	&	&
	\\
	rs7896554	&NACAP2	&0.840	&	&-0.630	&-0.565	&	&
	\\
	rs7897847	&LGI1	&	&	&	&	&	&0.962
	\\
	rs870753	&CFAP58	&	&	&	&-0.783	&	&
	\\
	rs881726	&GFRA1	&	&	&	&	&	&1.001
	\\
\end{longtable}

\begin{longtable}{ l l c c c  c c c}
	\caption[Analysis of the NHS T2D data using RBL-SS.]{Analysis of the NHS T2D data using RBL-SS.} \label{T2D.RBL.SS} \\
	\hline
	SNP & $\text{Gene}^\ast$ &    & chol & act  & gl    & ceraf & alcohol \\ \hline 
	\endfirsthead
	\caption[]{Continued from the previous page.} \\ \hline
	SNP & $\text{Gene}^\ast$ &    & chol & act  & gl    & ceraf & alcohol \\ \hline 
	\endhead
	\hline 
	\multicolumn{8}{r@{}}{Continued on the next page}\\
	\endfoot
	\hline
	\multicolumn{8}{l@{}}{$\ast$ Genes that SNPs belong to or are the closest to.}\\
	\endlastfoot
	& &  &1.354 &-0.430 &-0.778 &-2.424 &-4.000
	\\
	rs1041168	&PLPP4	&0.463	&	&	&	&	&
	\\
	rs10741150	&DOCK1	&	&-1.126	&	&	&	&
	\\
	rs10786611	&RF00019	&0.632	&	&	&	&	&
	\\
	rs10794069	&ADAM12	&0.524	&	&	&	&	&
	\\
	rs10824802	&MBL2	&0.553	&	&	&	&	&
	\\
	rs10884466	&RNA5SP326	&-0.439	&	&0.503	&	&	&
	\\
	rs10885423	&NRG3	&	&	&	&-1.060	&	&
	\\
	rs10886047	&MIR3663HG	&	&	&	&	&	&-0.410
	\\
	rs10886442	&GRK5	&	&1.087	&	&	&	&
	\\
	rs10998780	&ATP5MC1P7	&	&	&	&0.150	&	&
	\\
	rs11003665	&RNA5SP318	&	&0.632	&	&	&	&
	\\
	rs11013740	&KIAA1217	&	&	&	&	&	&0.852
	\\
	rs11196539	&NRG3	&	&	&	&	&	&-0.624
	\\
	rs11198590	&CACUL1	&-0.628	&	&	&	&	&
	\\
	rs11202221	&BMPR1A	&	&	&	&0.815	&	&
	\\
	rs11259039	&FRMD4A	&1.021	&	&	&	&	&
	\\
	rs11595123	&AKR1E2	&	&	&1.079	&	&	&
	\\
	rs11813505	&KIAA1217	&	&	&	&	&1.301	&
	\\
	rs1194657	&THAP12P3	&	&	&	&0.663	&	&
	\\
	rs1219508	&RPS15AP5	&-0.886	&	&	&	&	&
	\\
	rs12265854	&SLC16A12	&0.596	&	&	&	&	&
	\\
	rs12269237	&RF00017	&	&	&	&	&0.884	&
	\\
	rs12414552	&TCERG1L	&0.594	&	&	&	&	&
	\\
	rs12414627	&PNLIPRP1	&	&	&	&-0.551	&	&
	\\
	rs12767723	&SLC25A18P1	&0.962	&	&	&	&	&
	\\
	rs12772559	&TACR2	&	&	&	&0.906	&	&
	\\
	rs12774333	&LRMDA	&-0.449	&	&	&	&	&
	\\
	rs12775160	&FOXI2	&-0.560	&	&	&	&	&
	\\
	rs1573137	&SORCS3	&	&	&	&0.615	&	&
	\\
	rs16916794	&SLC39A12	&-0.803	&0.528	&	&	&	&
	\\
	rs16920092	&PLXDC2	&	&	&	&	&	&-0.655
	\\
	rs17094114	&GFRA1	&-0.563	&	&	&	&	&
	\\
	rs2291314	&PLPP4	&0.536	&	&	&	&	&
	\\
	rs2420979	&TACC2	&	&	&	&	&-1.091	&
	\\
	rs2492664	&OR6L1P	&0.655	&	&	&-0.363	&	&
	\\
	rs2664339	&RNU6-543P	&-0.501	&	&	&	&	&
	\\
	rs2666236	&IATPR	&	&	&	&	&0.689	&
	\\
	rs2784767	&PLAC9	&-0.452	&0.730	&	&	&	&
	\\
	rs2814322	&GRID1	&	&	&	&-0.806	&	&
	\\
	rs2842129	&DYNC1I2P1	&-0.662	&	&	&	&	&
	\\
	rs2900814	&SNRPD2P1	&-0.643	&	&	&	&	&
	\\
	rs3740063	&ABCC2	&	&	&	&-0.885	&	&
	\\
	rs3763722	&LARP4B	&	&	&	&	&	&1.036
	\\
	rs4411238	&PRKG1	&0.399	&	&	&	&	&
	\\
	rs4578341	&CHST15	&	&-0.582	&	&	&0.479	&
	\\
	rs4747009	&LRRC20	&	&	&	&	&	&0.710
	\\
	rs4747517	&ITIH5	&	&-0.905	&	&	&	&
	\\
	rs4749926	&IL2RA	&-0.607	&	&	&	&	&
	\\
	rs4752432	&PLPP4	&	&0.725	&	&	&	&
	\\
	rs4917817	&PYROXD2	&-0.506	&	&	&	&	&
	\\
	rs4934762	&PCAT5	&-0.560	&	&	&	&	&
	\\
	rs6482836	&DOCK1	&-0.709	&	&	&	&	&
	\\
	rs7070789	&GPAM	&	&	&	&-0.820	&	&
	\\
	rs7072255	&ANTXRLP1	&	&0.811	&	&	&	&
	\\
	rs7077721	&SNRPD2P1	&0.702	&	&	&	&	&
	\\
	rs7894809	&PCGF5	&0.501	&	&	&	&	&
	\\
	rs7896554	&NACAP2	&0.850	&	&-0.953	&	&	&
	\\
	rs7897847	&LGI1	&	&	&	&	&	&0.929
	\\
	rs7903853	&FRMD4A	&	&-1.185	&	&	&	&
	\\
	rs7920351	&TCERG1L	&	&	&-0.713	&	&	&
	\\
	rs881726	&GFRA1	&	&	&	&	&	&0.675
	\\
	rs943213	&DOCK1	&	&	&-0.939	&	&	&
	\\
\end{longtable} 

\begin{longtable}{ l l c c c  c c c}
	\caption[Analysis of the NHS T2D data using BSG-SS.]{Analysis of the NHS T2D data using BSG-SS.} \label{T2D.BSG.SS} \\
	\hline
	SNP & $\text{Gene}^\ast$ &    & chol & act  & gl    & ceraf & alcohol \\ \hline 
	\endfirsthead
	\caption[]{Continued from the previous page.} \\ \hline
	SNP & $\text{Gene}^\ast$ &    & chol & act  & gl    & ceraf & alcohol \\ \hline 
	\endhead
	\hline 
	\multicolumn{8}{r@{}}{Continued on the next page}\\
	\endfoot
	\hline
	\multicolumn{8}{l@{}}{$\ast$ Genes that SNPs belong to or are the closest to.}\\
	\endlastfoot
	& &  &2.045 &-2.049 &-2.204 &-1.796 &-4.436
	\\
	rs1041168	&PLPP4	&0.638	&	&	&	&	&
	\\
	rs10765059	&TCERG1L	&	&	&0.709	&	&	&
	\\
	rs10786611	&RF00019	&0.773	&0.556	&	&	&	&
	\\
	rs10829671	&EBF3	&-0.505	&	&	&	&	&
	\\
	rs10884466	&RNA5SP326	&-0.563	&	&0.625	&	&	&
	\\
	rs10886442	&GRK5	&	&1.038	&	&	&	&
	\\
	rs10998780	&ATP5MC1P7	&	&	&	&0.704	&	&
	\\
	rs11017821	&TCERG1L	&	&	&	&	&	&0.665
	\\
	rs11198590	&CACUL1	&-0.698	&	&	&0.905	&0.568	&
	\\
	rs11200996	&CCSER2	&	&	&	&	&0.508	&
	\\
	rs11259039	&FRMD4A	&1.174	&	&	&	&	&
	\\
	rs1219508	&RPS15AP5	&-0.787	&	&	&	&	&
	\\
	rs12265854	&SLC16A12	&0.681	&	&	&-0.494	&	&
	\\
	rs12269237	&RF00017	&	&	&	&	&0.684	&
	\\
	rs12414552	&TCERG1L	&0.480	&	&	&	&	&
	\\
	rs12764378	&ARID5B	&-0.420	&	&	&	&	&
	\\
	rs12767723	&SLC25A18P1	&0.638	&	&	&	&-0.559	&
	\\
	rs12775160	&FOXI2	&-0.762	&	&	&	&	&0.893
	\\
	rs1361709	&PCDH15	&	&	&-0.852	&	&	&
	\\
	rs1395465	&RN7SL63P	&0.292	&	&	&	&	&
	\\
	rs16916794	&SLC39A12	&-0.614	&0.580	&0.622	&	&	&
	\\
	rs16920092	&PLXDC2	&	&	&	&	&	&-0.692
	\\
	rs17094114	&GFRA1	&-0.676	&	&	&	&	&
	\\
	rs17469499	&KIAA1217	&	&	&	&	&-0.527	&
	\\
	rs2472737	&RET	&0.629	&	&	&	&	&
	\\
	rs2577356	&GFRA1	&	&	&	&	&0.875	&
	\\
	rs2784767	&PLAC9	&-0.569	&0.537	&	&	&	&
	\\
	rs2792708	&GPAM	&0.488	&	&	&	&	&
	\\
	rs2900814	&SNRPD2P1	&-0.460	&	&	&	&	&
	\\
	rs2926458	&RNU6-463P	&-0.680	&	&	&	&	&
	\\
	rs3763722	&LARP4B	&	&-1.251	&	&	&	&1.186
	\\
	rs4411238	&PRKG1	&0.619	&	&	&	&	&
	\\
	rs4747517	&ITIH5	&	&-1.257	&	&	&	&
	\\
	rs4752432	&PLPP4	&	&0.956	&	&	&	&
	\\
	rs4917817	&PYROXD2	&-0.626	&	&0.630	&	&	&
	\\
	rs4922535	&GDF10	&	&	&-0.601	&	&-0.649	&
	\\
	rs4934762	&PCAT5	&-0.640	&	&	&	&	&
	\\
	rs4934858	&NRP1	&0.281	&	&	&	&	&
	\\
	rs6482836	&DOCK1	&-0.773	&	&	&	&	&
	\\
	rs7070789	&GPAM	&	&	&	&-0.642	&	&
	\\
	rs7072255	&ANTXRLP1	&	&0.723	&	&	&	&
	\\
	rs7085788	&RHOBTB1	&-0.720	&	&	&	&	&
	\\
	rs7086058	&RN7SKP143	&-0.507	&	&	&	&	&
	\\
	rs716168	&VTI1A	&	&-0.570	&	&	&	&
	\\
	rs7894809	&PCGF5	&0.642	&	&	&	&	&
	\\
	rs7895870	&RN7SKP167	&	&-0.867	&	&	&	&
	\\
	rs7896554	&NACAP2	&1.097	&	&-0.477	&	&	&
	\\
	rs7917422	&HTR7	&	&	&	&	&	&0.794
	\\
	rs881726	&GFRA1	&	&	&	&	&	&0.933
	\\
\end{longtable} 

\begin{longtable}{ l l c c c  c c c}
	\caption[Analysis of the NHS T2D data using BL-SS.]{Analysis of the NHS T2D data using BL-SS.} \label{T2D.BL.SS} \\
	\hline
	SNP & $\text{Gene}^\ast$ &    & chol & act  & gl    & ceraf & alcohol \\ \hline 
	\endfirsthead
	\caption[]{Continued from the previous page.} \\ \hline
	SNP & $\text{Gene}^\ast$ &    & chol & act  & gl    & ceraf & alcohol \\ \hline 
	\endhead
	\hline 
	\multicolumn{8}{r@{}}{Continued on the next page}\\
	\endfoot
	\hline
	\multicolumn{8}{l@{}}{$\ast$ Genes that SNPs belong to or are the closest to.}\\
	\endlastfoot
	& &  &3.095 &-2.406 &-2.373 &-1.716 &-3.721
	\\
	rs1041168	&PLPP4	&0.670	&	&	&	&	&
	\\
	rs10508670	&KIAA1217	&	&0.773	&	&	&	&
	\\
	rs10765059	&TCERG1L	&	&	&0.445	&	&	&
	\\
	rs10829671	&EBF3	&-0.717	&	&	&	&	&
	\\
	rs10884466	&RNA5SP326	&-0.528	&	&	&	&	&
	\\
	rs10998780	&ATP5MC1P7	&	&	&	&1.195	&	&
	\\
	rs11017821	&TCERG1L	&	&	&	&	&	&0.307
	\\
	rs11198590	&CACUL1	&	&	&	&1.273	&	&
	\\
	rs11200996	&CCSER2	&	&	&	&	&0.509	&
	\\
	rs11202221	&BMPR1A	&	&	&	&0.954	&	&
	\\
	rs11259039	&FRMD4A	&1.020	&	&	&	&	&
	\\
	rs11594070	&ATE1-AS1	&	&	&	&-0.401	&	&
	\\
	rs1194657	&THAP12P3	&	&	&	&0.681	&	&
	\\
	rs12248205	&CDH23	&	&	&	&-0.938	&	&
	\\
	rs12256982	&ZMIZ1	&	&	&	&0.152	&	&
	\\
	rs12265854	&SLC16A12	&0.661	&	&	&-0.830	&	&
	\\
	rs12269237	&RF00017	&	&	&	&	&0.864	&
	\\
	rs12412976	&RPLP1P10	&	&	&0.592	&-0.590	&	&
	\\
	rs12414552	&TCERG1L	&0.549	&	&	&	&	&
	\\
	rs12414627	&PNLIPRP1	&	&	&	&-0.572	&	&
	\\
	rs12764378	&ARID5B	&-0.564	&	&	&	&	&
	\\
	rs12767723	&SLC25A18P1	&1.062	&	&	&	&	&
	\\
	rs12775160	&FOXI2	&-0.636	&	&	&	&	&
	\\
	rs1361709	&PCDH15	&	&	&-0.729	&	&	&
	\\
	rs1395465	&RN7SL63P	&0.562	&	&	&	&	&
	\\
	rs1573137	&SORCS3	&	&	&	&0.869	&	&
	\\
	rs16916794	&SLC39A12	&-0.430	&0.862	&	&	&	&
	\\
	rs16920092	&PLXDC2	&	&	&	&	&	&-0.508
	\\
	rs17094114	&GFRA1	&-0.734	&	&	&	&	&
	\\
	rs17469499	&KIAA1217	&	&	&	&	&-0.680	&
	\\
	rs2384105	&SNRPEP8	&	&	&	&	&	&-0.738
	\\
	rs2420979	&TACC2	&-0.629	&	&	&	&	&
	\\
	rs2472737	&RET	&0.553	&	&	&	&	&
	\\
	rs2577356	&GFRA1	&	&	&	&	&0.739	&
	\\
	rs2784767	&PLAC9	&-0.593	&0.576	&	&	&	&
	\\
	rs2792708	&GPAM	&0.568	&	&	&	&	&
	\\
	rs2900814	&SNRPD2P1	&	&	&	&	&-0.157	&
	\\
	rs2926458	&RNU6-463P	&-0.527	&	&	&	&	&
	\\
	rs3763722	&LARP4B	&	&-1.002	&	&	&	&1.151
	\\
	rs4411238	&PRKG1	&0.461	&	&	&	&	&
	\\
	rs4747009	&LRRC20	&	&	&	&	&	&1.016
	\\
	rs4747517	&ITIH5	&	&-1.695	&	&	&	&
	\\
	rs4752432	&PLPP4	&	&	&	&	&	&0.787
	\\
	rs4917817	&PYROXD2	&-0.637	&	&0.751	&	&	&
	\\
	rs4934762	&PCAT5	&-0.771	&	&	&	&	&
	\\
	rs4934858	&NRP1	&0.496	&	&	&	&	&
	\\
	rs6482836	&DOCK1	&-0.899	&	&	&	&	&
	\\
	rs7069001	&WDFY4	&	&-0.942	&	&	&	&
	\\
	rs7070789	&GPAM	&	&	&	&-1.154	&	&-0.771
	\\
	rs7077718	&DNMBP	&-0.661	&	&	&	&	&
	\\
	rs7085788	&RHOBTB1	&-0.721	&	&	&	&	&
	\\
	rs7086058	&RN7SKP143	&-0.872	&	&	&	&	&
	\\
	rs716168	&VTI1A	&	&-0.662	&	&	&	&
	\\
	rs7894809	&PCGF5	&0.828	&	&	&	&	&
	\\
	rs7895870	&RN7SKP167	&	&-1.295	&	&	&	&
	\\
	rs7896554	&NACAP2	&0.989	&	&	&	&	&
	\\
	rs7917422	&HTR7	&	&	&	&0.663	&	&1.306
	\\
	rs7920351	&TCERG1L	&	&	&	&-1.059	&	&
	\\
	rs809836	&LYZL1	&	&	&	&1.109	&	&
	\\
	rs881726	&GFRA1	&	&	&	&	&	&0.922
	\\
	rs915216	&DUSP5	&	&1.102	&	&	&	&
	\\
\end{longtable}


\begin{longtable}{ l c c c c c}
	\caption{Analysis of the TCGA SKCM data using RBSG-SS.} \label{SKCM.RBSG.SS} \\
	\hline
	Gene &    & clark & stage  & age   & gender \\ \hline 
	\endfirsthead
	\caption[]{Continued from the previous page.} \\ \hline
	Gene &    & clark & stage  & age   & gender \\ \hline 
	\endhead
	\hline 
	\multicolumn{6}{r@{}}{Continued on the next page}\\
	\endfoot
	\hline
	\endlastfoot
	& & 0.834  &0.228 &-0.116 &-0.183
	\\ 
	AHNAKRS	&0.107	&	&	&	&
	\\
	ANKRD28	&0.134	&	&0.138	&	&
	\\
	ASH2L	&	&-0.297	&	&	&
	\\
	BTD	&	&-0.312	&	&	&
	\\
	C1ORF140	&-0.002	&0.246	&-0.083	&-0.022	&0.092
	\\
	CD44	&	&	&	&	&0.070
	\\
	CHP1	&0.107	&0.045	&	&	&
	\\
	CXCL6	&0.126	&-0.120	&-0.095	&	&
	\\
	DLG6	&0.113	&-0.015	&0.067	&0.185	&-0.142
	\\
	DOK5	&	&	&	&-0.066	&
	\\
	ETNK2	&0.152	&	&	&	&
	\\
	FILIP1	&-0.030	&	&	&	&
	\\
	JADE1	&-0.147	&	&	&	&
	\\
	JPH4	&	&0.115	&	&	&
	\\
	KBF2	&-0.032	&0.182	&	&0.034	&-0.026
	\\
	LRRN2	&-0.061	&	&	&	&
	\\
	MAGED4	&-0.098	&	&-0.020	&	&
	\\
	NHSL2	&-0.088	&	&	&	&
	\\
	PITPNA	&0.151	&-0.051	&-0.012	&-0.033	&0.008
	\\
	SOX8	&0.088	&	&-0.212	&	&
	\\
	TMEM145	&	&	&	&	&0.048
	\\
	TMEM159	&0.160	&-0.121	&-0.042	&	&0.189
	\\
	WBSCR27	&	&0.070	&	&0.126	&
	\\
\end{longtable}

\begin{longtable}{ l c c c c c}
	\caption{Analysis of the TCGA SKCM data using RBL-SS.} \label{SKCM.RBL.SS} \\
	\hline
	Gene &    & clark & stage  & age   & gender \\ \hline 
	\endfirsthead
	\caption[]{Continued from the previous page.} \\ \hline
	Gene &    & clark & stage  & age   & gender \\ \hline 
	\endhead
	\hline 
	\multicolumn{6}{r@{}}{Continued on the next page}\\
	\endfoot
	\hline
	\endlastfoot
	&&0.926 &-0.062 &-0.011  &0.388
	\\
	AHNAKRS	&0.084	&	&	&	&
	\\
	ANKRD28	&0.191	&	&0.207	&	&
	\\
	ASH2L	&	&-0.258	&	&	&
	\\
	BAIAP2	&0.043	&	&	&	&
	\\
	BTD	&	&-0.309	&	&-0.255	&
	\\
	C1ORF140	&	&0.129	&	&	&
	\\
	C1ORF54	&	&	&	&	&-0.102
	\\
	CHP1	&0.081	&	&	&	&-0.111
	\\
	CPXM1	&	&	&	&0.005	&
	\\
	CSNK2A2	&-0.003	&	&	&	&
	\\
	CYP1B1-AS1	&	&	&0.104	&	&
	\\
	DAP	&0.036	&	&	&-0.116	&
	\\
	DLG6	&	&	&	&0.242	&
	\\
	ETNK2	&0.109	&	&	&	&
	\\
	FHL5	&	&0.220	&	&	&
	\\
	FILIP1	&	&	&-0.016	&	&
	\\
	GAMT	&	&	&	&0.082	&
	\\
	IL11RA	&-0.087	&	&	&	&
	\\
	IQCK	&	&	&	&	&-0.090
	\\
	JADE1	&-0.161	&	&	&	&
	\\
	JPH4	&	&0.159	&	&	&
	\\
	KDM6B	&	&	&-0.142	&	&
	\\
	LRFN2	&	&0.096	&	&	&
	\\
	MAGED4	&-0.130	&	&	&	&
	\\
	MAPE	&	&	&	&	&-0.191
	\\
	MPD1	&-0.078	&	&	&	&
	\\
	NHSL2	&-0.144	&-0.306	&	&	&
	\\
	PAX1	&0.171	&	&0.217	&	&
	\\
	PBX2	&0.141	&	&	&0.130	&
	\\
	PITPNA	&0.161	&	&	&-0.056	&
	\\
	RNPEPL1	&	&	&0.052	&	&
	\\
	SLC12A5	&	&	&	&	&-0.081
	\\
	SOX8	&0.140	&	&-0.091	&	&
	\\
	STPG1	&	&	&0.184	&	&
	\\
	TMEM145	&	&	&	&	&0.222
	\\
	TMEM159	&0.123	&	&	&	&
	\\
	TNFAIP1	&	&	&	&0.283	&
	\\
	TP53TG1	&	&0.102	&	&	&-0.063
	\\
	WBSCR27	&	&0.090	&	&0.126	&
	\\
\end{longtable}

\begin{longtable}{ l c c c c c}
	\caption{Analysis of the TCGA SKCM data using BSG-SS.} \label{SKCM.BSG.SS} \\
	\hline
	Gene &    & clark & stage  & age   & gender \\ \hline 
	\endfirsthead
	\caption[]{Continued from the previous page.} \\ \hline
	Gene &    & clark & stage  & age   & gender \\ \hline 
	\endhead
	\hline 
	\multicolumn{6}{r@{}}{Continued on the next page}\\
	\endfoot
	\hline
	\endlastfoot
	& &0.487 &0.163 &0.048 &0.087
	\\
	AHNAKRS	&0.120	&	&	&	&
	\\
	ANKRD28	&0.138	&	&	&	&
	\\
	ARMC9	&0.008	&	&	&	&
	\\
	ASH2L	&0.019	&-0.194	&	&-0.107	&
	\\
	BTD	&	&-0.303	&	&-0.138	&
	\\
	C14ORF2	&	&0.251	&	&	&
	\\
	C1ORF140	&	&0.100	&0.024	&	&0.029
	\\
	CD44	&	&	&	&	&0.125
	\\
	CHP1	&0.123	&	&	&	&
	\\
	CPXM1	&-0.047	&	&	&	&
	\\
	CXCL6	&0.032	&	&	&	&
	\\
	DLG6	&	&0.093	&	&0.204	&-0.061
	\\
	DOK5	&	&	&-0.052	&	&
	\\
	ETNK2	&0.094	&	&	&	&
	\\
	FILIP1	&-0.049	&	&	&	&
	\\
	GAMT	&	&-0.004	&	&	&
	\\
	IL11RA	&-0.045	&	&	&	&
	\\
	JADE1	&-0.149	&	&	&	&
	\\
	JPH4	&	&0.110	&	&	&
	\\
	KBF2	&-0.077	&	&	&	&
	\\
	LRRN2	&-0.073	&	&	&	&
	\\
	MAGED4	&-0.122	&	&	&	&
	\\
	MAPE	&	&	&	&	&-0.217
	\\
	NHSL2	&-0.026	&	&	&	&
	\\
	PBX2	&0.133	&	&	&0.155	&
	\\
	PHP1B	&-0.076	&	&	&	&
	\\
	PITPNA	&0.150	&0.077	&	&0.038	&-0.039
	\\
	SOX8	&0.103	&	&-0.148	&	&
	\\
	STPG1	&	&	&0.197	&	&
	\\
	TMEM145	&0.015	&-0.045	&	&	&0.147
	\\
	TMEM159	&0.140	&	&	&	&0.113
	\\
	TNFRSF4	&	&0.077	&	&	&
	\\
	TP53TG1	&	&0.072	&	&	&
	\\
	WBSCR27	&	&0.015	&	&0.092	&
	\\
	ZFP62	&-0.010	&	&	&	&
	\\
\end{longtable}

\begin{longtable}{ l c c c c c}
	\caption{Analysis of the TCGA SKCM data using BL-SS.} \label{SKCM.BL.SS} \\
	\hline
	Gene &    & clark & stage  & age   & gender \\ \hline 
	\endfirsthead
	\caption[]{Continued from the previous page.} \\ \hline
	Gene &    & clark & stage  & age   & gender \\ \hline 
	\endhead
	\hline 
	\multicolumn{6}{r@{}}{Continued on the next page}\\
	\endfoot
	\hline
	\endlastfoot
	& &0.545 &0.308 &0.080 &0.047
	\\
	AHNAKRS	&0.102	&	&	&	&
	\\
	ANKRD28	&0.180	&	&0.134	&	&
	\\
	ASH2L	&	&	&	&-0.185	&
	\\
	BTD	&	&-0.386	&	&	&
	\\
	C14ORF2	&	&0.126	&	&	&
	\\
	C1ORF140	&	&0.199	&	&	&
	\\
	CELSR2	&	&	&0.112	&	&
	\\
	CHP1	&0.080	&	&	&	&
	\\
	CPXM1	&-0.067	&	&	&	&
	\\
	CSNK2A2	&-0.026	&	&	&	&
	\\
	CYP1B1-AS1	&	&	&0.104	&	&
	\\
	DAP	&	&	&	&-0.139	&
	\\
	DLG6	&0.088	&	&	&0.236	&
	\\
	ETNK2	&0.206	&	&	&	&-0.089
	\\
	FHL5	&	&0.076	&	&	&
	\\
	FILIP1	&	&	&-0.062	&	&
	\\
	GAMT	&	&	&	&0.058	&
	\\
	IL11RA	&-0.056	&	&	&	&
	\\
	IQCK	&	&	&	&	&-0.098
	\\
	JADE1	&-0.203	&	&	&	&
	\\
	JPH4	&	&0.101	&	&	&
	\\
	KBF2	&-0.089	&	&	&	&
	\\
	KDM6B	&	&	&-0.173	&	&
	\\
	LRFN2	&	&0.109	&	&	&
	\\
	LRRN2	&-0.091	&	&	&	&
	\\
	MAGED4	&-0.113	&	&	&	&
	\\
	MAPE	&	&	&	&	&-0.114
	\\
	MPD1	&-0.100	&	&	&	&
	\\
	NHSL2	&-0.035	&	&	&	&
	\\
	PAX1	&	&	&0.050	&	&
	\\
	PBX2	&0.126	&	&	&0.072	&
	\\
	PHP1B	&	&	&	&	&-0.054
	\\
	PIP4K2C	&	&	&-0.101	&	&
	\\
	PITPNA	&0.193	&	&	&	&
	\\
	PTP4A3	&	&	&-0.138	&	&
	\\
	RNPEPL1	&	&	&0.171	&	&
	\\
	SAA2	&0.021	&	&	&-0.058	&
	\\
	SLC12A5	&	&	&	&	&-0.112
	\\
	SOX8	&0.132	&	&-0.084	&	&
	\\
	TIE1	&-0.093	&	&	&	&
	\\
	TMEM145	&	&	&	&	&0.188
	\\
	TMEM159	&0.174	&	&	&	&0.181
	\\
	TP53TG1	&	&0.156	&	&	&-0.030
	\\
	WBSCR27	&	&0.048	&	&0.105	&
	\\
\end{longtable}

\section{Biological similarity analysis}
\begin{figure}[h!]
	\centering
	\includegraphics[angle=0,origin=c,width=1\textwidth]{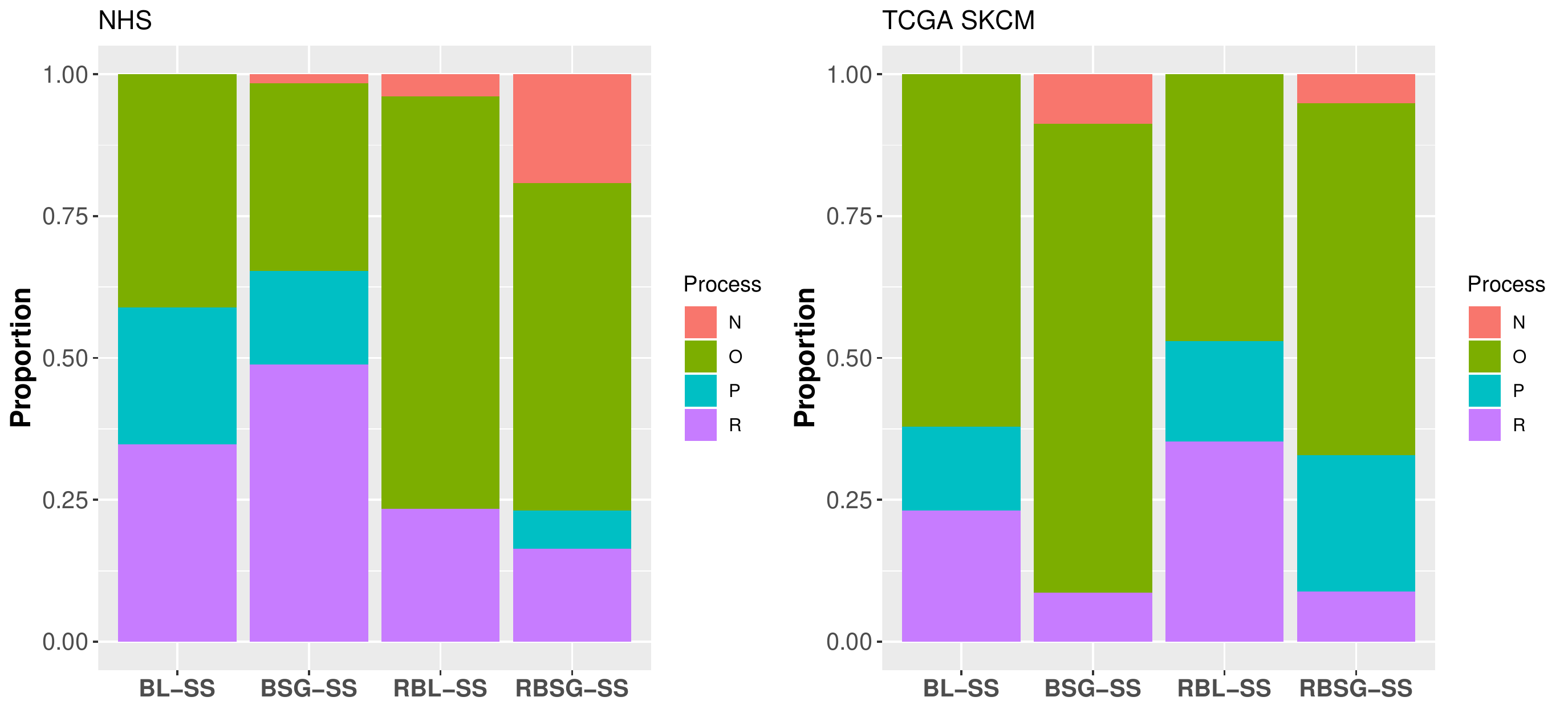}
	\caption[Gene Ontology analysis]{Gene Ontology (GO) analysis: proportions of genes that have the four categories of processes with different approaches. Left: NHS data. Right: TCGA SKCM data.}
	\label{fig:go}
\end{figure}

We carried out an examination of the Gene Ontology (GO) biological processes which provide us with a deeper insight on the differences of the markers identified by different methods.  We totally identified 77 unique genes using our proposed method along with three other methods for the NHS data.  We conducted the GO enrichment analysis using the R package GOSim and found these genes involve in a total of 158 GO biological processes, the p-values of which are smaller than 0.1 in the GO enrichment analysis.  Then we divided the 158 processes into four categories: positive regulation (P), negative regulation (N), regulation (R, without a well-defined “direction”) and other (O).
We computed the proportions of genes that involve in the four categories of processes for each of the four methods. 
Similarly, for the TCGA SKCM data, 109 genes were identified by our method along with three other alternative methods. GO enrichment analysis showed that they involve in 183 biological processes, with p-values smaller than 0.1.
The results for NHS and TCGA SKCM are provided in Figure \ref{fig:go}, which shows an obvious difference between our proposed method and the three alternatives in both datasets.

\section{Posterior inference}

\subsection{RBG-SS}
\subsubsection{Hierarchical model specification}

\setlength{\jot}{5pt}
\begin{gather*}
y_{i} = W_{i}^\top\alpha + E_{i}^\top\theta + U_{i}^\top\beta + \nu^{-\frac{1}{2}}\kappa \sqrt{u_{i}}z_{i} \quad i=1,\dots,n\\
u_{i}|\nu \stackrel{ind}{\thicksim} \nu\exp\left(-\nu u_{i}\right) \quad i=1,\dots,n\\
z_{i} \stackrel{ind}{\thicksim} \text{N}(0,\; 1) \quad i=1,\dots,n\\ 
\nu \thicksim \text{Gamma}\left(c_{1}, \; c_{2}\right) \\
\alpha \thicksim \text{N}_{q}(0, \, \Sigma_{\alpha0}) \\
\theta \thicksim \text{N}_{k}(0, \, \Sigma_{\theta0}) \\
\beta_{j}|\phi_{j}, s_{j} \stackrel{ind}{\thicksim} \phi_{j}\, \text{N}_{L}\left(0,\, s_{j}\textbf{I}_{L}\right) + (1-\phi_{j})\delta_{0}(\beta_{j}) \quad j=1,\dots,p\\
\phi_{j}|\pi_{0} \stackrel{ind}{\thicksim} \text{Bernoulli}(\pi_{0}) \quad j=1,\dots,p\\
\pi_{0} \thicksim \text{Beta}\left(a_{0}, \, b_{0}\right) \\
s_{j}|\eta \thicksim \text{Gamma}\left(\frac{L+1}{2}, \frac{\eta}{2}\right) \quad j=1,\dots,p\\
\eta \thicksim \text{Gamma}\left(d_{1},\; d_{2}\right)\\
\end{gather*}

\subsubsection{Gibbs Sampler}
\begin{itemize}
	\item $u_{i}|\text{rest} \thicksim \text{Inverse-Gaussian}(\mu_{u_{i}}, \, \lambda_{u_{i}})$, where the shape parameter $\lambda_{u_{i}}=2\nu$, mean parameter $\mu_{u_{i}}=\sqrt{\frac{2\kappa^{2}}{(y_{i}-\tilde{y_{i}})^{2}}}$ and $\tilde{y_{i}}=y_{i}-W_{i}^\top\alpha-E_{i}^\top\theta-U_{i}^\top\beta$.
	
	\item $\nu|\text{rest} \thicksim \text{Gamma}\left(s_{\nu}, \, r_{\nu} \right)$, where the shape parameter $s_{\nu}=c_{1}+\frac{3n}{2}$ and the rate parameter $r_{\nu}=c_{2}+\sum_{i=1}^{n}u_{i}+(2\kappa^{2})^{-1}\sum_{i=1}^{n}u_{i}^{-1}\tilde{y_{i}}^{2}$. 
	
	\item $\alpha|\text{rest} \thicksim \text{N}(\mu_{\alpha}, \, \Sigma_{\alpha})$, where
	\begin{equation*}
	\setlength{\jot}{10pt}
	\begin{aligned}
	\mu_{\alpha} &= \Sigma_{\alpha}\nu\kappa^{-2}\sum_{i=1}^{n}u_{i}^{-1}W_{i}(y_{i}-E_{i}^\top\theta-U_{i}^\top\beta) \\
	\Sigma_{\alpha} &= \left(\nu\kappa^{-2}\sum_{i=1}^{n}u_{i}^{-1}W_{i}W_{i}^\top + \Sigma_{\alpha0}^{-1} \right)^{-1} 
	\end{aligned}
	\end{equation*}
	
	\item $\theta|\text{rest} \thicksim \text{N}(\mu_{\theta}, \, \Sigma_{\theta})$, where
	\begin{equation*}
	\setlength{\jot}{10pt}
	\begin{aligned}
	\mu_{\theta} &= \Sigma_{\theta}\nu\kappa^{-2}\sum_{i=1}^{n}u_{i}^{-1}E_{i}(y_{i}-W_{i}^\top\alpha-U_{i}^\top\beta) \\
	\Sigma_{\theta} &= \left(\nu\kappa^{-2}\sum_{i=1}^{n}u_{i}^{-1}E_{i}E_{i}^\top + \Sigma_{\theta0}^{-1} \right)^{-1} 
	\end{aligned}
	\end{equation*}
	
	\item $\beta_{j}|\text{rest} \thicksim l_{j} \text{N}(\mu_{\beta_{j}}, \, \Sigma_{\beta_{j}}) + (1-l_{j})\delta_{0}(\beta_{j})$ where
	\begin{equation*}
	\setlength{\jot}{10pt}
	\begin{aligned}
	\mu_{\beta_{j}} &= \Sigma_{\beta_{j}}\nu\kappa^{-2}\sum_{i=1}^{n}u_{i}^{-1}U_{ij}\tilde{y}_{ij} \\
	\Sigma_{\beta_{j}} &= \left(\nu\kappa^{-2}\sum_{i=1}^{n}u_{i}^{-1}U_{ij}U_{ij}^\top + \frac{1}{s_{j}}\textbf{I}_{L}\right)^{-1} \\
	l_{j} &= \frac{\pi_{0}}{\pi_{0} + (1-\pi_{0})s_{j}^{\frac{L}{2}}|\Sigma_{\beta_{j}}|^{-\frac{1}{2}} \exp\left\{- \frac{1}{2} \lVert\Sigma_{\beta_{j}}^{\frac{1}{2}}\nu\kappa^{-2}\sum_{i=1}^{n}u_{i}^{-1}U_{ij}\tilde{y}_{ij}\rVert_{2}^{2} \right\}}
	\end{aligned}
	\end{equation*}
	and $\tilde{y}_{ij}$ is defined as $\tilde{y}_{ij}=y_{i}-W_{i}^\top\alpha-E_{i}^\top\theta-U_{(j)}^\top\beta_{(j)}$.
	
	\item The posterior of $s_{j}$ is
	\begin{equation*}
	s_{j}^{-1}|\text{rest} \thicksim \begin{cases}
	\scalebox{1}{Inverse-Gamma($\frac{L+1}{2}$,\, $\frac{\eta}{2}$)}& { \text{if} \; \beta_{j} = 0} \\[5pt]
	\scalebox{1}{Inverse-Gaussian($\eta$, $\sqrt{\frac{\eta}{\lVert\beta_{j}\rVert_{2}^{2}}}$)}& { \text{if} \; \beta_{j} \neq 0}
	\end{cases}
	\end{equation*}
	
	\item $\pi_{0}|\text{rest} \thicksim \text{Beta}\left(a_{0}+\sum_{j=1}^{p}\textbf{I}_{\{\beta_{j} \neq 0\}}, \; b_{0}+\sum_{j=1}^{p}\textbf{I}_{\{\beta_{j} = 0\}}\right)$
	
	\item $\eta|\text{rest} \thicksim \text{Gamma}\left(s_{\eta}, \, r_{\eta} \right)$, where $s_{\eta}=\frac{p+p\times L}{2}+d_{1}$ and the rate parameter $r_{\eta}=\frac{\sum_{j=1}^{p}s_{j}}{2} + d_{2}$. 
	
\end{itemize}

\subsection{RBL-SS}
\subsubsection{Hierarchical model specification}

\setlength{\jot}{5pt}
\begin{gather*}
y_{i} = W_{i}^\top\alpha + E_{i}^\top\theta + U_{i}^\top\beta + \nu^{-\frac{1}{2}}\kappa \sqrt{u_{i}}z_{i} \\
u_{i}|\nu \stackrel{ind}{\thicksim} \nu\exp\left(-\nu u_{i}\right) \\
z_{i} \stackrel{ind}{\thicksim} \text{N}(0,\; 1) \\ 
\nu \thicksim \text{Gamma}\left(c_{1}, \; c_{2}\right) \\
\alpha \thicksim \text{N}_{q}(0, \, \Sigma_{\alpha0}) \\
\theta \thicksim \text{N}_{k}(0, \, \Sigma_{\theta0}) \\
\beta_{jl}|\phi_{jl}, s_{jl} \stackrel{ind}{\thicksim} \phi_{jl}\, \text{N} \left(0,\, s_{jl}\right) + (1-\phi_{jl})\delta_{0}(\beta_{jl}) \quad j=1,\dots,p;\; l=1,\dots,L \\
\phi_{jl}|\pi_{1} \stackrel{ind}{\thicksim} \text{Bernoulli}(\pi_{1}) \quad j=1,\dots,p;\; l=1,\dots,L\\
s_{jl}|\eta \thicksim \text{Gamma}\left(1, \frac{\eta}{2}\right) \quad j=1,\dots,p;\;  l=1,\dots,L\\
\pi_{1} \thicksim \text{Beta}\left(a_{1}, \, b_{1}\right) \\
\eta \thicksim \text{Gamma}\left(d_{1},\; d_{2}\right)\\
\end{gather*}
\subsubsection{Gibbs Sampler}
\begin{itemize}
	\item $u_{i}|\text{rest} \thicksim \text{Inverse-Gaussian}(\mu_{u_{i}}, \, \lambda_{u_{i}})$, where the shape parameter $\lambda_{u_{i}}=2\nu$, mean parameter $\mu_{u_{i}}=\sqrt{\frac{2\kappa^{2}}{(Y_{i}-\tilde{y_{i}})^{2}}}$ and $\tilde{y_{i}}=y_{i}-W_{i}^\top\alpha-E_{i}^\top\theta-U_{i}^\top\beta$.
	
	\item $\nu|\text{rest} \thicksim \text{Gamma}\left(s_{\nu}, \, r_{\nu} \right)$, where the shape parameter $s_{\nu}=c_{1}+\frac{3n}{2}$ and the rate parameter $r_{\nu}=c_{2}+\sum_{i=1}^{n}u_{i}+(2\kappa^{2})^{-1}\sum_{i=1}^{n}u_{i}^{-1}\tilde{y_{i}}^{2}$. 
	
	\item $\alpha|\text{rest} \thicksim \text{N}(\mu_{\alpha}, \, \Sigma_{\alpha})$, where
	\begin{equation*}
	\setlength{\jot}{10pt}
	\begin{aligned}
	\mu_{\alpha} &= \Sigma_{\alpha}\nu\kappa^{-2}\sum_{i=1}^{n}u_{i}^{-1}W_{i}(Y_{i}-E_{i}^\top\theta-U_{i}^\top\beta) \\
	\Sigma_{\alpha} &= \left(\nu\kappa^{-2}\sum_{i=1}^{n}u_{i}^{-1}W_{i}W_{i}^\top + \Sigma_{\alpha0}^{-1} \right)^{-1} 
	\end{aligned}
	\end{equation*}
	
	\item $\theta|\text{rest} \thicksim \text{N}(\mu_{\theta}, \, \Sigma_{\theta})$, where
	\begin{equation*}
	\setlength{\jot}{10pt}
	\begin{aligned}
	\mu_{\theta} &= \Sigma_{\theta}\nu\kappa^{-2}\sum_{i=1}^{n}u_{i}^{-1}E_{i}(Y_{i}-W_{i}^\top\alpha-U_{i}^\top\beta) \\
	\Sigma_{\theta} &= \left(\nu\kappa^{-2}\sum_{i=1}^{n}u_{i}^{-1}E_{i}E_{i}^\top + \Sigma_{\theta0}^{-1} \right)^{-1} 
	\end{aligned}
	\end{equation*}
	
	\item $\beta_{jl}|\text{rest} \thicksim l_{jl} \text{N}(\mu_{\beta_{jl}}, \, \sigma^{2}_{\beta_{jl}}) + (1-l_{jl})\delta_{0}(\beta_{jl})$ where
	\begin{equation*}
	\setlength{\jot}{10pt}
	\begin{aligned}
	\mu_{\beta_{jl}} &= \sigma_{\beta_{jl}}^{2}\nu\kappa^{-2}\sum_{i=1}^{n}u_{i}^{-1}U_{ijl}\tilde{y}_{ijl} \\
	\sigma_{\beta_{jl}}^{2} &= \left(\nu\kappa^{-2}\sum_{i=1}^{n}u_{i}^{-1}U_{ijl}^{2} + \frac{1}{s_{jl}}\right)^{-1} \\
	l_{jl} &= \frac{\pi_{1}}{\pi_{1} + (1-\pi_{1})s_{jl}^{\frac{1}{2}}(\sigma_{\beta_{jl}}^{2})^{-\frac{1}{2}} \exp\left\{- \frac{1}{2} \sigma_{\beta_{jl}}^{2}(\nu\kappa^{-2}\sum_{i=1}^{n}u_{i}^{-1}U_{ijl}\tilde{y}_{ijl})^{2} \right\}}
	\end{aligned}
	\end{equation*}
	and $\tilde{y}_{ijl}$ is defined as $\tilde{y}_{ijl}=y_{i}-W_{i}^\top\alpha-E_{i}^\top\theta-U_{(jl)}^\top\beta_{(jl)}$.
	
	\item The posterior of $s_{jl}$ is
	\begin{equation*}
	s_{jl}^{-1}|\text{rest} \thicksim \begin{cases}
	\scalebox{1}{Inverse-Gamma(1,\, $\frac{\eta}{2}$)}& { \text{if} \; \beta_{jl} = 0} \\[5pt]
	\scalebox{1}{Inverse-Gaussian($\eta$, $\sqrt{\frac{\eta}{\beta_{jl}^{2}}}$)}& { \text{if} \; \beta_{jl} \neq 0}
	\end{cases}
	\end{equation*}
	
	\item $\pi_{1}|\text{rest} \thicksim \text{Beta}\left(a_{1}+\sum_{j,l}\textbf{I}_{\{\beta_{jl} \neq 0\}}, \; b_{1}+\sum_{j,l}\textbf{I}_{\{\beta_{jl} = 0\}}\right)$
	
	\item $\eta|\text{rest} \thicksim \text{Gamma}\left(s_{\eta}, \, r_{\eta} \right)$, where $s_{\eta}=p\times L+d_{1}$ and the rate parameter $r_{\eta}=\frac{\sum_{j,l}s_{jl}}{2} + d_{2}$. 
\end{itemize}


\subsection{RBSG}
\subsubsection{Hierarchical model specification}

\setlength{\jot}{7pt}
\begin{gather*}
y_{i} = W_{i}^\top\alpha + E_{i}^\top\theta + U_{i}^\top\beta + \nu^{-\frac{1}{2}}\kappa \sqrt{u_{i}}z_{i} \\
u_{i}|\nu \stackrel{ind}{\thicksim} \nu\exp\left(-\nu u_{i}\right) \\
z_{i} \stackrel{ind}{\thicksim} \text{N}(0,\; 1) \\ 
\nu \thicksim \text{Gamma}\left(c_{1}, \; c_{2}\right) \\
\alpha \thicksim \text{N}_{q}(0, \, \Sigma_{\alpha0}) \\
\theta \thicksim \text{N}_{k}(0, \, \Sigma_{\theta0}) \\
\beta_{j}| r_{j}, \omega_{jl} \stackrel{ind}{\thicksim} \text{N}_{L}(0, \, V_{j}), \;\text{where}\; V_{j}=\text{diag}\bigg\{\big(\frac{1}{r_{j}}+\frac{1}{\omega_{jl}}\big)^{-1},\; l=1,2,\dots,L\bigg\}\\ 
r_{j}, \omega_{j1}, \dots, \omega_{jL}|\eta_{1}, \eta_{2} \propto \prod_{l=1}^{L} \left[(\omega_{jl})^{-\frac{1}{2}} \left(\frac{1}{r_{j}}+\frac{1}{\omega_{jl}}\right)^{-\frac{1}{2}}\right] (r_{j})^{-\frac{1}{2}} \exp\left(-\frac{\eta_{1}}{2}r_{j}-\frac{\eta_{2}}{2}\sum_{l=1}^{L}\omega_{jl}\right)\\
\eta_{1}, \eta_{2} \propto \eta_{1}^{\frac{p}{2}}\eta_{2}^{pL} \exp\left\{-d_{1}\eta_{1}-d_{2}\eta_{2}\right\}\\
\sigma^{2} \thicksim 1/\sigma^{2}
\end{gather*}

\subsubsection{Gibbs Sampler}

\begin{itemize}
	
	\item $u_{i}|\text{rest} \thicksim \text{Inverse-Gaussian}(\mu_{u_{i}}, \, \lambda_{u_{i}})$, where the shape parameter $\lambda_{u_{i}}=2\nu$, mean parameter $\mu_{u_{i}}=\sqrt{\frac{2\kappa^{2}}{(Y_{i}-\tilde{y_{i}})^{2}}}$ and $\tilde{y_{i}}=y_{i}-W_{i}^\top\alpha-E_{i}^\top\theta-U_{i}^\top\beta$.
	
	\item $\nu|\text{rest} \thicksim \text{Gamma}\left(s_{\nu}, \, r_{\nu} \right)$, where the shape parameter $s_{\nu}=c_{1}+\frac{3n}{2}$ and the rate parameter $r_{\nu}=c_{2}+\sum_{i=1}^{n}u_{i}+(2\kappa^{2})^{-1}\sum_{i=1}^{n}u_{i}^{-1}\tilde{y_{i}}^{2}$. 
	
	\item $\alpha|\text{rest} \thicksim \text{N}(\mu_{\alpha}, \, \Sigma_{\alpha})$, where
	\begin{equation*}
	\setlength{\jot}{10pt}
	\begin{aligned}
	\mu_{\alpha} &= \Sigma_{\alpha}\nu\kappa^{-2}\sum_{i=1}^{n}u_{i}^{-1}W_{i}(Y_{i}-E_{i}^\top\theta-U_{i}^\top\beta) \\
	\Sigma_{\alpha} &= \left(\nu\kappa^{-2}\sum_{i=1}^{n}u_{i}^{-1}W_{i}W_{i}^\top + \Sigma_{\alpha0}^{-1} \right)^{-1} 
	\end{aligned}
	\end{equation*}
	
	\item $\theta|\text{rest} \thicksim \text{N}(\mu_{\theta}, \, \Sigma_{\theta})$, where
	\begin{equation*}
	\setlength{\jot}{10pt}
	\begin{aligned}
	\mu_{\theta} &= \Sigma_{\theta}\nu\kappa^{-2}\sum_{i=1}^{n}u_{i}^{-1}E_{i}(Y_{i}-W_{i}^\top\alpha-U_{i}^\top\beta) \\
	\Sigma_{\theta} &= \left(\nu\kappa^{-2}\sum_{i=1}^{n}u_{i}^{-1}E_{i}E_{i}^\top + \Sigma_{\theta0}^{-1} \right)^{-1} 
	\end{aligned}
	\end{equation*}
	
	\item $\beta_{j}|\text{rest} \thicksim \text{N}(\mu_{\beta_{j}}, \, \Sigma_{\beta_{j}})$ where
	\begin{equation*}
	\setlength{\jot}{10pt}
	\begin{aligned}
	\mu_{\beta_{j}} &= \Sigma_{\beta_{j}}\nu\kappa^{-2}\sum_{i=1}^{n}u_{i}^{-1}U_{ij}\tilde{y}_{ij} \\
	\Sigma_{\beta_{j}} &= \left(\nu\kappa^{-2}\sum_{i=1}^{n}u_{i}^{-1}U_{ij}U_{ij}^\top + V_{j}^{-1}\right)^{-1}
	\end{aligned}
	\end{equation*}
	and $\tilde{y}_{ij}$ is defined as $\tilde{y}_{ij}=y_{i}-W_{i}^\top\alpha-E_{i}^\top\theta-U_{(j)}^\top\beta_{(j)}$.
	
	\item $r_{j}^{-1}|\text{rest} \thicksim \text{Inv-Gaussian}(\eta_{1}, \, \sqrt{\frac{\eta_{1}\sigma^{2}}{\lVert\beta_{j}\rVert_{2}^{2}}})$
	
	\item $\omega_{jl}^{-1}|\text{rest} \thicksim \text{Inv-Gaussian}(\eta_{2}, \, \sqrt{\frac{\eta_{2}\sigma^{2}}{\beta_{jl}^{2}}})$
	
	\item $\eta_{1}|\text{rest} \thicksim \text{Gamma}\left(s_{\eta_{1}}, \, r_{\eta_{1}} \right)$, where $s_{\eta_{1}}=\frac{p}{2}+1$ and the rate parameter $r_{\eta_{1}}=\frac{\sum_{j=1}^{p}r_{j}}{2} + d_{1}$. 
	
	\item $\eta_{2}|\text{rest} \thicksim \text{Gamma}\left(s_{\eta_{2}}, \, r_{\eta_{2}} \right)$, where $s_{\eta_{2}}=p\times L+1$ and the rate parameter $r_{\eta_{2}}=\frac{\sum_{j,l}\omega_{jl}}{2} + d_{2}$. 
	
\end{itemize}

\subsection{RBG}
\subsubsection{Hierarchical model specification}

\setlength{\jot}{5pt}
\begin{gather*}
y_{i} = W_{i}^\top\alpha + E_{i}^\top\theta + U_{i}^\top\beta + \nu^{-\frac{1}{2}}\kappa \sqrt{u_{i}}z_{i} \quad i=1,\dots,n\\
u_{i}|\nu \stackrel{ind}{\thicksim} \nu\exp\left(-\nu u_{i}\right) \quad i=1,\dots,n\\
z_{i} \stackrel{ind}{\thicksim} \text{N}(0,\; 1) \quad i=1,\dots,n\\ 
\nu \thicksim \text{Gamma}\left(c_{1}, \; c_{2}\right) \\
\alpha \thicksim \text{N}_{q}(0, \, \Sigma_{\alpha0}) \\
\theta \thicksim \text{N}_{k}(0, \, \Sigma_{\theta0}) \\
\beta_{j}|s_{j} \stackrel{ind}{\thicksim} \text{N}_{L}\left(0,\, s_{j}\textbf{I}_{L}\right) \quad j=1,\dots,p\\
s_{j}|\eta \thicksim \text{Gamma}\left(\frac{L+1}{2}, \frac{\eta}{2}\right) \quad j=1,\dots,p\\
\eta \thicksim \text{Gamma}\left(d_{1},\; d_{2}\right)\\
\end{gather*}

\subsubsection{Gibbs Sampler}
\begin{itemize}
	\item $u_{i}|\text{rest} \thicksim \text{Inverse-Gaussian}(\mu_{u_{i}}, \, \lambda_{u_{i}})$, where the shape parameter $\lambda_{u_{i}}=2\nu$, mean parameter $\mu_{u_{i}}=\sqrt{\frac{2\kappa^{2}}{(y_{i}-\tilde{y_{i}})^{2}}}$ and $\tilde{y_{i}}=y_{i}-W_{i}^\top\alpha-E_{i}^\top\theta-U_{i}^\top\beta$.
	
	\item $\nu|\text{rest} \thicksim \text{Gamma}\left(s_{\nu}, \, r_{\nu} \right)$, where the shape parameter $s_{\nu}=c_{1}+\frac{3n}{2}$ and the rate parameter $r_{\nu}=c_{2}+\sum_{i=1}^{n}u_{i}+(2\kappa^{2})^{-1}\sum_{i=1}^{n}u_{i}^{-1}\tilde{y_{i}}^{2}$. 
	
	\item $\alpha|\text{rest} \thicksim \text{N}(\mu_{\alpha}, \, \Sigma_{\alpha})$, where
	\begin{equation*}
	\setlength{\jot}{10pt}
	\begin{aligned}
	\mu_{\alpha} &= \Sigma_{\alpha}\nu\kappa^{-2}\sum_{i=1}^{n}u_{i}^{-1}W_{i}(y_{i}-E_{i}^\top\theta-U_{i}^\top\beta) \\
	\Sigma_{\alpha} &= \left(\nu\kappa^{-2}\sum_{i=1}^{n}u_{i}^{-1}W_{i}W_{i}^\top + \Sigma_{\alpha0}^{-1} \right)^{-1} 
	\end{aligned}
	\end{equation*}
	
	\item $\theta|\text{rest} \thicksim \text{N}(\mu_{\theta}, \, \Sigma_{\theta})$, where
	\begin{equation*}
	\setlength{\jot}{10pt}
	\begin{aligned}
	\mu_{\theta} &= \Sigma_{\theta}\nu\kappa^{-2}\sum_{i=1}^{n}u_{i}^{-1}E_{i}(y_{i}-W_{i}^\top\alpha-U_{i}^\top\beta) \\
	\Sigma_{\theta} &= \left(\nu\kappa^{-2}\sum_{i=1}^{n}u_{i}^{-1}E_{i}E_{i}^\top + \Sigma_{\theta0}^{-1} \right)^{-1} 
	\end{aligned}
	\end{equation*}
	
	\item $\beta_{j}|\text{rest} \thicksim  \text{N}(\mu_{\beta_{j}}, \, \Sigma_{\beta_{j}})$ where
	\begin{equation*}
	\setlength{\jot}{10pt}
	\begin{aligned}
	\mu_{\beta_{j}} &= \Sigma_{\beta_{j}}\nu\kappa^{-2}\sum_{i=1}^{n}u_{i}^{-1}U_{ij}\tilde{y}_{ij} \\
	\Sigma_{\beta_{j}} &= \left(\nu\kappa^{-2}\sum_{i=1}^{n}u_{i}^{-1}U_{ij}U_{ij}^\top + \frac{1}{s_{j}}\textbf{I}_{L}\right)^{-1}
	\end{aligned}
	\end{equation*}
	and $\tilde{y}_{ij}$ is defined as $\tilde{y}_{ij}=y_{i}-W_{i}^\top\alpha-E_{i}^\top\theta-U_{(j)}^\top\beta_{(j)}$.
	
	\item $s_{j}^{-1}|\text{rest} \thicksim \text{Inverse-Gaussian}(\eta, \, \sqrt{\frac{\eta}{\lVert\beta_{j}\rVert_{2}^{2}}})$
	
	\item $\eta|\text{rest} \thicksim \text{Gamma}\left(s_{\eta}, \, r_{\eta} \right)$, where $s_{\eta}=\frac{p+p\times L}{2}+d_{1}$ and the rate parameter $r_{\eta}=\frac{\sum_{j=1}^{p}s_{j}}{2} + d_{2}$. 
	
\end{itemize}

\subsection{RBL}
\subsubsection{Hierarchical model specification}

\setlength{\jot}{5pt}
\begin{gather*}
y_{i} = W_{i}^\top\alpha + E_{i}^\top\theta + U_{i}^\top\beta + \nu^{-\frac{1}{2}}\kappa \sqrt{u_{i}}z_{i} \\
u_{i}|\nu \stackrel{ind}{\thicksim} \nu\exp\left(-\nu u_{i}\right) \\
z_{i} \stackrel{ind}{\thicksim} \text{N}(0,\; 1) \\ 
\nu \thicksim \text{Gamma}\left(c_{1}, \; c_{2}\right) \\
\alpha \thicksim \text{N}_{q}(0, \, \Sigma_{\alpha0}) \\
\theta \thicksim \text{N}_{k}(0, \, \Sigma_{\theta0}) \\
\beta_{jl}|s_{jl} \stackrel{ind}{\thicksim}  \text{N} \left(0,\, s_{jl}\right)  \quad j=1,\dots,p;\; l=1,\dots,L \\
s_{jl}|\eta \thicksim \text{Gamma}\left(1, \frac{\eta}{2}\right) \quad j=1,\dots,p;\;  l=1,\dots,L \\
\eta \thicksim \text{Gamma}\left(d_{1},\; d_{2}\right)\\
\end{gather*}
\subsubsection{Gibbs Sampler}
\begin{itemize}
	\item $u_{i}|\text{rest} \thicksim \text{Inverse-Gaussian}(\mu_{u_{i}}, \, \lambda_{u_{i}})$, where the shape parameter $\lambda_{u_{i}}=2\nu$, mean parameter $\mu_{u_{i}}=\sqrt{\frac{2\kappa^{2}}{(Y_{i}-\tilde{y_{i}})^{2}}}$ and $\tilde{y_{i}}=y_{i}-W_{i}^\top\alpha-E_{i}^\top\theta-U_{i}^\top\beta$.
	
	\item $\nu|\text{rest} \thicksim \text{Gamma}\left(s_{\nu}, \, r_{\nu} \right)$, where the shape parameter $s_{\nu}=c_{1}+\frac{3n}{2}$ and the rate parameter $r_{\nu}=c_{2}+\sum_{i=1}^{n}u_{i}+(2\kappa^{2})^{-1}\sum_{i=1}^{n}u_{i}^{-1}\tilde{y_{i}}^{2}$. 
	
	\item $\alpha|\text{rest} \thicksim \text{N}(\mu_{\alpha}, \, \Sigma_{\alpha})$, where
	\begin{equation*}
	\setlength{\jot}{10pt}
	\begin{aligned}
	\mu_{\alpha} &= \Sigma_{\alpha}\nu\kappa^{-2}\sum_{i=1}^{n}u_{i}^{-1}W_{i}(Y_{i}-E_{i}^\top\theta-U_{i}^\top\beta) \\
	\Sigma_{\alpha} &= \left(\nu\kappa^{-2}\sum_{i=1}^{n}u_{i}^{-1}W_{i}W_{i}^\top + \Sigma_{\alpha0}^{-1} \right)^{-1} 
	\end{aligned}
	\end{equation*}
	
	\item $\theta|\text{rest} \thicksim \text{N}(\mu_{\theta}, \, \Sigma_{\theta})$, where
	\begin{equation*}
	\setlength{\jot}{10pt}
	\begin{aligned}
	\mu_{\theta} &= \Sigma_{\theta}\nu\kappa^{-2}\sum_{i=1}^{n}u_{i}^{-1}E_{i}(Y_{i}-W_{i}^\top\alpha-U_{i}^\top\beta) \\
	\Sigma_{\theta} &= \left(\nu\kappa^{-2}\sum_{i=1}^{n}u_{i}^{-1}E_{i}E_{i}^\top + \Sigma_{\theta0}^{-1} \right)^{-1} 
	\end{aligned}
	\end{equation*}
	
	\item $\beta_{jl}|\text{rest} \thicksim  \text{N}(\mu_{\beta_{jl}}, \, \sigma^{2}_{\beta_{jl}})$ where
	\begin{equation*}
	\setlength{\jot}{10pt}
	\begin{aligned}
	\mu_{\beta_{jl}} &= \sigma_{\beta_{jl}}^{2}\nu\kappa^{-2}\sum_{i=1}^{n}u_{i}^{-1}U_{ijl}\tilde{y}_{ijl} \\
	\sigma_{\beta_{jl}}^{2} &= \left(\nu\kappa^{-2}\sum_{i=1}^{n}u_{i}^{-1}U_{ijl}^{2} + \frac{1}{s_{jl}}\right)^{-1}
	\end{aligned}
	\end{equation*}
	and $\tilde{y}_{ijl}$ is defined as $\tilde{y}_{ijl}=y_{i}-W_{i}^\top\alpha-E_{i}^\top\theta-U_{(jl)}^\top\beta_{(jl)}$.
	
	\item $s_{j}^{-1}|\text{rest} \thicksim \text{Inverse-Gaussian}(\eta, \, \sqrt{\frac{\eta}{\beta_{jl}^{2}}})$
	
	\item $\eta|\text{rest} \thicksim \text{Gamma}\left(s_{\eta}, \, r_{\eta} \right)$, where $s_{\eta}=p\times L+d_{1}$ and the rate parameter $r_{\eta}=\frac{\sum_{j,l}s_{jl}}{2} + d_{2}$. 
\end{itemize}


\subsection{BSG-SS}
\subsubsection{Hierarchical model specification}

\setlength{\jot}{7pt}
\begin{gather*}
Y \propto (\sigma^{2})^{-\frac{n}{2}} \exp\left\{-\frac{1}{2\sigma^{2}} \sum_{i=1}^{n}(y_{i}- W_{i}^\top\alpha - E_{i}^\top\theta - U_{i}^\top\beta)^{2} \right\} \\ 
\alpha \thicksim \text{N}_{q}(0, \, \Sigma_{\alpha0}) \\
\theta \thicksim \text{N}_{k}(0, \, \Sigma_{\theta0}) \\
\beta_{j} = V_{j}^{\frac{1}{2}}b_{j}, \quad \text{where} \; V_{j}^{\frac{1}{2}} = \text{diag}\{\omega_{j1},\dots,\omega_{jL}\} \\
b_{j}|\phi^{b}_{j} \stackrel{ind}{\thicksim} \phi^{b}_{j}\, \text{N}_{L}\left(0,\, \textbf{I}_{L}\right) + (1-\phi^{b}_{j})\delta_{0}(b_{j}) \\
\phi^{b}_{j}|\pi_{0} \stackrel{ind}{\thicksim} \text{Bernoulli}(\pi_{0}) \\
\pi_{0} \thicksim \text{Beta}\left(a_{0}, \, b_{0}\right) \\
\omega_{jl}|\phi^{w}_{jl} \stackrel{ind}{\thicksim} \phi^{w}_{jl}\, \text{N}^{+}\left(0,\, s^{2}\right) + (1-\phi^{w}_{jl})\delta_{0}(\omega_{jl}) \\
\phi^{w}_{jl}|\pi_{1} \stackrel{ind}{\thicksim} \text{Bernoulli}(\pi_{1}) \\
\pi_{1} \thicksim \text{Beta}\left(a_{1}, \, b_{1}\right) \\
s^{2} \thicksim \text{Inverse-Gamma}\left(1, \eta\right)\\
\sigma^{2} \thicksim 1/\sigma^{2}
\end{gather*}

\subsubsection{Gibbs Sampler}

\begin{itemize}
	
	\item $\alpha|\text{rest} \thicksim \text{N}(\mu_{\alpha}, \, \Sigma_{\alpha})$, where
	\begin{equation*}
	\setlength{\jot}{10pt}
	\begin{aligned}
	\mu_{\alpha} &= \Sigma_{\alpha}(\sigma^{2})^{-1}\sum_{i=1}^{n}W_{i}(y_{i}-E_{i}^\top\theta-U_{i}^\top\beta) \\
	\Sigma_{\alpha} &= \left(\frac{1}{\sigma^{2}}\sum_{i=1}^{n}W_{i}W_{i}^\top + \Sigma_{\alpha0}^{-1} \right)^{-1} 
	\end{aligned}
	\end{equation*}
	
	\item $\theta|\text{rest} \thicksim \text{N}(\mu_{\theta}, \, \Sigma_{\theta})$, where
	\begin{equation*}
	\setlength{\jot}{10pt}
	\begin{aligned}
	\mu_{\theta} &= \Sigma_{\theta}(\sigma^{2})^{-1}\sum_{i=1}^{n}E_{i}(y_{i}-W_{i}^\top\alpha-U_{i}^\top\beta) \\
	\Sigma_{\theta} &= \left(\frac{1}{\sigma^{2}}\sum_{i=1}^{n}E_{i}E_{i}^\top + \Sigma_{\theta0}^{-1} \right)^{-1} 
	\end{aligned}
	\end{equation*}
	
	\item $b_{j}|\text{rest} \thicksim l_{j} \text{N}(\mu_{b_{j}}, \, \Sigma_{b_{j}}) + (1-l_{j})\delta_{0}(b_{j})$ where
	\begin{equation*}
	\setlength{\jot}{10pt}
	\begin{aligned}
	\mu_{b_{j}} &= \Sigma_{b_{j}}(\sigma^{2})^{-1}\sum_{i=1}^{n}V_{j}^{\frac{1}{2}}U_{ij}\tilde{y}_{ij} \\
	\Sigma_{b_{j}} &= \left(\frac{1}{\sigma^{2}} \sum_{i=1}^{n}V_{j}^{\frac{1}{2}}U_{ij}U_{ij}^\top V_{j}^{\frac{1}{2}} + \textbf{I}_{L}\right)^{-1} \\
	l^{b}_{j} &= \frac{\pi_{0}}{\pi_{0} + (1-\pi_{0})|\Sigma_{b_{j}}|^{-\frac{1}{2}} 
		\exp\left\{- \frac{1}{2\sigma^{4}} \lVert\Sigma_{b_{j}}^{\frac{1}{2}}\sum_{i=1}^{n}V_{j}^{\frac{1}{2}}U_{ij}\tilde{y}_{ij}\rVert_{2}^{2} \right\}}
	\end{aligned}
	\end{equation*}
	and $\tilde{y}_{ij}$ is defined as $\tilde{y}_{ij}=y_{i}-W_{i}^\top\alpha-E_{i}^\top\theta-U_{(j)}^\top\beta_{(j)}$.
	
	\item $\omega_{jl}|\text{rest} \thicksim l^{w}_{jl}\; \text{N}^{+}(\mu_{\omega_{jl}}, \, \sigma^{2}_{\omega_{jl}}) + (1-l^{w}_{jl})\delta_{0}(\omega_{jl})$ where
	\begin{equation*}
	\setlength{\jot}{10pt}
	\begin{aligned}
	\mu_{\omega_{jl}} &= \sigma_{\omega_{jl}}^{2}(\sigma^{2})^{-1}\sum_{i=1}^{n}b_{jl}U_{ijl}\tilde{y}_{ijl} \\
	\sigma_{\omega_{jl}}^{2} &= \left(\frac{1}{\sigma^{2}} \sum_{i=1}^{n}U_{ijl}^{2} b_{jl}^{2} + \frac{1}{s^{2}}\right)^{-1} \\
	l^{w}_{jl} &= \frac{\pi_{1}}{\pi_{1} + (1-\pi_{1})\frac{1}{2}s(\sigma_{\omega_{jl}}^{2})^{-\frac{1}{2}} \exp\left\{- \frac{\sigma_{\omega_{jl}}^{2}}{2\sigma^{4}} \left(\sum_{i=1}^{n}b_{jl}U_{ijl}\tilde{y}_{ijl}\right)^{2} \right\}\left[\Phi\left(\frac{\mu_{\omega_{jl}}}{\sigma_{\omega_{jl}}}\right)\right]^{-1}} 
	\end{aligned}
	\end{equation*}
	and $\tilde{y}_{ijl}$ is defined as $\tilde{y}_{ijl}=y_{i}-W_{i}^\top\alpha-E_{i}^\top\theta-U_{(jl)}^\top\beta_{(jl)}$.
	
	\item $s^{2}|\text{rest} \thicksim \text{Inv-Gamma}\left(1+\frac{1}{2}\sum_{j,l}\textbf{I}_{\{\omega_{jl} \neq 0\}},\; \eta+\frac{1}{2}\sum_{j,l}\omega_{jl}^{2}\right)$
	
	\item $\pi_{0}|\text{rest} \thicksim \text{Beta}\left(a_{0}+\sum_{j=1}^{p}\textbf{I}_{\{\beta_{j} \neq 0\}}, \; b_{0}+\sum_{j=1}^{p}\textbf{I}_{\{\beta_{j} = 0\}}\right)$
	
	\item $\pi_{1}|\text{rest} \thicksim \text{Beta}\left(a_{1}+\sum_{j,l}\textbf{I}_{\{\omega_{jl} \neq 0\}}, \; b_{1}+\sum_{j,l}\textbf{I}_{\{\omega_{jl} = 0\}}\right)$
	
	\item $\eta$ is estimated with the EM approach used in the proposed method. For the gth EM update $\eta^{(g)} = \frac{1}{E_{\eta^{(g-1)}} \left[\frac{1}{s^{2}}|Y\right]}$.
	
	\item $\sigma^{2}|\text{rest} \thicksim \text{Inv-Gamma}(\frac{n}{2},\; \frac{\sum_{i=1}^{n}\tilde{y_{i}}^{2}}{2})$, where $\tilde{y_{i}}=y_{i}-W_{i}^\top\alpha-E_{i}^\top\theta-U_{i}^\top\beta$.
	
\end{itemize}

\subsection{BGL-SS}
\subsubsection{Hierarchical model specification}

\setlength{\jot}{7pt}
\begin{gather*}
Y \propto (\sigma^{2})^{-\frac{n}{2}} \exp\left\{-\frac{1}{2\sigma^{2}} \sum_{i=1}^{n}(y_{i}- W_{i}^\top\alpha - E_{i}^\top\theta - U_{i}^\top\beta)^{2} \right\} \\ 
\alpha \thicksim \text{N}_{q}(0, \, \Sigma_{\alpha0}) \\
\theta \thicksim \text{N}_{k}(0, \, \Sigma_{\theta0}) \\
\beta_{j}|\phi_{j}, \sigma^{2}, s_{j} \stackrel{ind}{\thicksim} \phi_{j}\, \text{N}_{L}\left(0,\, \sigma^{2}s_{j}\textbf{I}_{L}\right) + (1-\phi_{j})\delta_{0}(\beta_{j}) \quad j=1,\dots,p\\
\phi_{j}|\pi_{0} \stackrel{ind}{\thicksim} \text{Bernoulli}(\pi_{0}) \quad j=1,\dots,p\\
\pi_{0} \thicksim \text{Beta}\left(a_{0}, \, b_{0}\right) \\
s_{j}|\eta \stackrel{ind}{\thicksim} \text{Gamma}\left(\frac{L+1}{2}, \frac{\eta}{2}\right) \quad j=1,\dots,p\\
\eta \thicksim \text{Gamma}\left(d_{1},\; d_{2}\right)\\
\sigma^{2} \thicksim 1/\sigma^{2}
\end{gather*}

\subsubsection{Gibbs Sampler}

\begin{itemize}
	
	\item $\alpha|\text{rest} \thicksim \text{N}(\mu_{\alpha}, \, \Sigma_{\alpha})$, where
	\begin{equation*}
	\setlength{\jot}{10pt}
	\begin{aligned}
	\mu_{\alpha} &= \Sigma_{\alpha}(\sigma^{2})^{-1}\sum_{i=1}^{n}W_{i}(y_{i}-E_{i}^\top\theta-U_{i}^\top\beta) \\
	\Sigma_{\alpha} &= \left(\frac{1}{\sigma^{2}}\sum_{i=1}^{n}W_{i}W_{i}^\top + \Sigma_{\alpha0}^{-1} \right)^{-1} 
	\end{aligned}
	\end{equation*}
	
	\item $\theta|\text{rest} \thicksim \text{N}(\mu_{\theta}, \, \Sigma_{\theta})$, where
	\begin{equation*}
	\setlength{\jot}{10pt}
	\begin{aligned}
	\mu_{\theta} &= \Sigma_{\theta}(\sigma^{2})^{-1}\sum_{i=1}^{n}E_{i}(y_{i}-W_{i}^\top\alpha-U_{i}^\top\beta) \\
	\Sigma_{\theta} &= \left(\frac{1}{\sigma^{2}}\sum_{i=1}^{n}E_{i}E_{i}^\top + \Sigma_{\theta0}^{-1} \right)^{-1} 
	\end{aligned}
	\end{equation*}
	
	\item $\beta_{j}|\text{rest} \thicksim l_{j} \text{N}(\mu_{\beta_{j}}, \, \sigma^{2}\Sigma_{\beta_{j}}) + (1-l_{j})\delta_{0}(\beta_{j})$ where
	\begin{equation*}
	\setlength{\jot}{10pt}
	\begin{aligned}
	\mu_{\beta_{j}} &= \Sigma_{\beta_{j}}\sum_{i=1}^{n}U_{ij}\tilde{y}_{ij} \\
	\Sigma_{\beta_{j}} &= \left(\sum_{i=1}^{n}U_{ij}U_{ij}^\top + \frac{1}{s_{j}}\textbf{I}_{L} \right)^{-1} \\
	l_{j} &= \frac{\pi_{0}}{\pi_{0} + (1-\pi_{0})s_{j}^{\frac{L}{2}}|\Sigma_{\beta_{j}}|^{-\frac{1}{2}} \exp\left\{- \frac{1}{2\sigma^{2}} \lVert\Sigma_{\beta_{j}}^{\frac{1}{2}}\sum_{i=1}^{n}U_{ij}\tilde{y}_{ij}\rVert_{2}^{2} \right\}}
	\end{aligned}
	\end{equation*}
	and $\tilde{y}_{ij}$ is defined as $\tilde{y}_{ij}=y_{i}-W_{i}^\top\alpha-E_{i}^\top\theta-U_{(j)}^\top\beta_{(j)}$.
	
	\item The posterior of $s_{j}$ is
	\begin{equation*}
	s_{j}^{-1}|\text{rest} \thicksim \begin{cases}
	\scalebox{1}{Inverse-Gamma($\frac{L+1}{2}$,\, $\frac{\eta}{2}$)}& { \text{if} \; \beta_{j} = 0} \\[5pt]
	\scalebox{1}{Inverse-Gaussian($\eta$, $\sqrt{\frac{\eta \sigma^{2}}{\lVert\beta_{j}\rVert_{2}^{2}}}$)}& { \text{if} \; \beta_{j} \neq 0}
	\end{cases}
	\end{equation*}
	
	\item $\pi_{0}|\text{rest} \thicksim \text{Beta}\left(a_{0}+\sum_{j=1}^{p}\textbf{I}_{\{\beta_{j} \neq 0\}}, \; b_{0}+\sum_{j=1}^{p}\textbf{I}_{\{\beta_{j} = 0\}}\right)$
	
	\item $\eta|\text{rest} \thicksim \text{Gamma}\left(s_{\eta}, \, r_{\eta} \right)$, where $s_{\eta}=\frac{p+p\times L}{2}+d_{1}$ and the rate parameter $r_{\eta}=\frac{\sum_{j=1}^{p}s_{j}}{2} + d_{2}$. 
	
	\item $\sigma^{2}|\text{rest} \thicksim \text{Inv-Gamma}(\frac{n+L\sum_{j=1}^{p}\textbf{I}_{\{\beta_{j}\neq0\}}}{2} ,\; \frac{\sum_{i=1}^{n}\tilde{y_{i}}^{2} + \sum_{j=1}^{p}(s_{j})^{-1}\beta_{j}^\top \beta_{j} }{2})$, where $\tilde{y_{i}}=y_{i}-W_{i}^\top\alpha-E_{i}^\top\theta-U_{i}^\top\beta$.
	
\end{itemize}

\subsection{BL-SS}
\subsubsection{Hierarchical model specification}

\setlength{\jot}{7pt}
\begin{gather*}
Y \propto (\sigma^{2})^{-\frac{n}{2}} \exp\left\{-\frac{1}{2\sigma^{2}} \sum_{i=1}^{n}(y_{i}- W_{i}^\top\alpha - E_{i}^\top\theta - U_{i}^\top\beta)^{2} \right\} \\ 
\alpha \thicksim \text{N}_{q}(0, \, \Sigma_{\alpha0}) \\
\theta \thicksim \text{N}_{k}(0, \, \Sigma_{\theta0}) \\
\beta_{jl}|\phi_{jl}, \sigma^{2}, s_{jl} \stackrel{ind}{\thicksim} \phi_{jl}\, \text{N} \left(0,\, \sigma^{2} s_{jl}\right) + (1-\phi_{jl})\delta_{0}(\beta_{jl}) \quad j=1,\dots,p;\; l=1,\dots,L \\
\phi_{jl}|\pi_{1} \stackrel{ind}{\thicksim} \text{Bernoulli}(\pi_{1}) \quad j=1,\dots,p;\; l=1,\dots,L\\
s_{jl}|\eta \stackrel{ind}{\thicksim} \text{Gamma}\left(1, \frac{\eta}{2}\right) \quad j=1,\dots,p;\;  l=1,\dots,L\\
\pi_{1} \thicksim \text{Beta}\left(a_{1}, \, b_{1}\right) \\
\eta \thicksim \text{Gamma}\left(d_{1},\; d_{2}\right)\\
\sigma^{2} \thicksim 1/\sigma^{2}
\end{gather*}

\subsubsection{Gibbs Sampler}

\begin{itemize}
	
	\item $\alpha|\text{rest} \thicksim \text{N}(\mu_{\alpha}, \, \Sigma_{\alpha})$, where
	\begin{equation*}
	\setlength{\jot}{10pt}
	\begin{aligned}
	\mu_{\alpha} &= \Sigma_{\alpha}(\sigma^{2})^{-1}\sum_{i=1}^{n}W_{i}(y_{i}-E_{i}^\top\theta-U_{i}^\top\beta) \\
	\Sigma_{\alpha} &= \left(\frac{1}{\sigma^{2}}\sum_{i=1}^{n}W_{i}W_{i}^\top + \Sigma_{\alpha0}^{-1} \right)^{-1} 
	\end{aligned}
	\end{equation*}
	
	\item $\theta|\text{rest} \thicksim \text{N}(\mu_{\theta}, \, \Sigma_{\theta})$, where
	\begin{equation*}
	\setlength{\jot}{10pt}
	\begin{aligned}
	\mu_{\theta} &= \Sigma_{\theta}(\sigma^{2})^{-1}\sum_{i=1}^{n}E_{i}(y_{i}-W_{i}^\top\alpha-U_{i}^\top\beta) \\
	\Sigma_{\theta} &= \left(\frac{1}{\sigma^{2}}\sum_{i=1}^{n}E_{i}E_{i}^\top + \Sigma_{\theta0}^{-1} \right)^{-1} 
	\end{aligned}
	\end{equation*}
	
	\item $\beta_{jl}|\text{rest} \thicksim l_{jl} \text{N}(\mu_{\beta_{jl}}, \, \sigma^{2}_{\beta_{jl}}) + (1-l_{jl})\delta_{0}(\beta_{jl})$ where
	\begin{equation*}
	\setlength{\jot}{10pt}
	\begin{aligned}
	\mu_{\beta_{jl}} &= \sigma^{2}_{\beta_{jl}}(\sigma^{2})^{-1}\sum_{i=1}^{n}U_{ijl}\tilde{y}_{ijl} \\
	\sigma^{2}_{\beta_{jl}} &= \sigma^{2}\left(\sum_{i=1}^{n}U_{ijl}^{2} + \frac{1}{s_{jl}} \right)^{-1} \\
	l_{jl} &= \frac{\pi_{0}}{\pi_{0} + (1-\pi_{0})s_{jl}^{\frac{1}{2}}(\sigma^{2}_{\beta_{jl}})^{-\frac{1}{2}}(\sigma^{2})^{-\frac{1}{2}} \exp\left\{- \frac{\sigma^{2}_{\beta_{jl}}}{2\sigma^{4}} \left(\sum_{i=1}^{n}U_{ijl}\tilde{y}_{ijl}\right)^{2} \right\}}
	\end{aligned}
	\end{equation*}
	and $\tilde{y}_{ijl}$ is defined as $\tilde{y}_{ijl}=y_{i}-W_{i}^\top\alpha-E_{i}^\top\theta-U_{(jl)}^\top\beta_{(jl)}$.

	\item The posterior of $s_{jl}$ is
	\begin{equation*}
	s_{jl}^{-1}|\text{rest} \thicksim \begin{cases}
	\scalebox{1}{Inverse-Gamma($1$,\, $\frac{\eta}{2}$)}& { \text{if} \; \beta_{jl} = 0} \\[5pt]
	\scalebox{1}{Inverse-Gaussian($\eta$, $\sqrt{\frac{\eta \sigma^{2}}{\beta_{jl}^{2}}}$)}& { \text{if} \; \beta_{jl} \neq 0}
	\end{cases}
	\end{equation*}
	
	\item $\pi_{1}|\text{rest} \thicksim \text{Beta}\left(a_{1}+\sum_{j,l}\textbf{I}_{\{\beta_{jl} \neq 0\}}, \; b_{1}+\sum_{j,l}\textbf{I}_{\{\beta_{jl} = 0\}}\right)$
	
	\item $\eta|\text{rest} \thicksim \text{Gamma}\left(s_{\eta}, \, r_{\eta} \right)$, where $s_{\eta}=p\times L+d_{1}$ and the rate parameter $r_{\eta}=\frac{\sum_{j,l}s_{jl}}{2} + d_{2}$. 
	
	\item $\sigma^{2}|\text{rest} \thicksim \text{Inv-Gamma}(\frac{n+\sum_{j,l}\textbf{I}_{\{\beta_{jl} \neq 0\}}}{2},\; \frac{\sum_{i=1}^{n}\tilde{y_{i}}^{2}+\sum_{j,l}(s_{jl}^{-1})\beta_{jl}^{2}}{2})$, where $\tilde{y_{i}}=y_{i}-W_{i}^\top\alpha-E_{i}^\top\theta-U_{i}^\top\beta$.
	
\end{itemize}


\subsection{BSG}
\subsubsection{Hierarchical model specification}

\setlength{\jot}{7pt}
\begin{gather*}
Y|\alpha,\theta,\beta,\sigma^{2} \propto (\sigma^{2})^{-\frac{n}{2}} \exp\left\{-\frac{1}{2\sigma^{2}} \sum_{i=1}^{n}(y_{i}- W_{i}^\top\alpha - E_{i}^\top\theta - U_{i}^\top\beta)^{2} \right\} \\ 
\alpha \thicksim \text{N}_{q}(0, \, \Sigma_{\alpha0}) \\
\theta \thicksim \text{N}_{k}(0, \, \Sigma_{\theta0}) \\
\beta_{j}|\omega_{jl}, r_{j} \stackrel{ind}{\thicksim} \text{N}_{L}(0, \, \sigma^{2}V_{j}), \;\text{where}\; V_{j}=\text{diag}\bigg\{\big(\frac{1}{r_{j}}+\frac{1}{\omega_{jl}}\big)^{-1},\; l=1,2,\dots,L\bigg\}\\ 
r_{j}, \omega_{j1}, \dots, \omega_{jL}|\eta_{1}, \eta_{2} \propto \prod_{l=1}^{L} \left[(\omega_{jl})^{-\frac{1}{2}} \left(\frac{1}{r_{j}}+\frac{1}{\omega_{jl}}\right)^{-\frac{1}{2}}\right] (r_{j})^{-\frac{1}{2}} \exp\left(-\frac{\eta_{1}}{2}r_{j}-\frac{\eta_{2}}{2}\sum_{l=1}^{L}\omega_{jl}\right)\\
\eta_{1}, \eta_{2} \propto \eta_{1}^{\frac{p}{2}}\eta_{2}^{pL} \exp\left\{-d_{1}\eta_{1}-d_{2}\eta_{2}\right\}\\
\sigma^{2} \thicksim 1/\sigma^{2}
\end{gather*}

\subsubsection{Gibbs Sampler}

\begin{itemize}
	
	\item $\alpha|\text{rest} \thicksim \text{N}(\mu_{\alpha}, \, \Sigma_{\alpha})$, where
	\begin{equation*}
	\setlength{\jot}{10pt}
	\begin{aligned}
	\mu_{\alpha} &= \Sigma_{\alpha}(\sigma^{2})^{-1}\sum_{i=1}^{n}W_{i}(y_{i}-E_{i}^\top\theta-U_{i}^\top\beta) \\
	\Sigma_{\alpha} &= \left(\frac{1}{\sigma^{2}}\sum_{i=1}^{n}W_{i}W_{i}^\top + \Sigma_{\alpha0}^{-1} \right)^{-1} 
	\end{aligned}
	\end{equation*}
	
	\item $\theta|\text{rest} \thicksim \text{N}(\mu_{\theta}, \, \Sigma_{\theta})$, where
	\begin{equation*}
	\setlength{\jot}{10pt}
	\begin{aligned}
	\mu_{\theta} &= \Sigma_{\theta}(\sigma^{2})^{-1}\sum_{i=1}^{n}E_{i}(y_{i}-W_{i}^\top\alpha-U_{i}^\top\beta) \\
	\Sigma_{\theta} &= \left(\frac{1}{\sigma^{2}}\sum_{i=1}^{n}E_{i}E_{i}^\top + \Sigma_{\theta0}^{-1} \right)^{-1} 
	\end{aligned}
	\end{equation*}
	
	\item $\beta_{j}|\text{rest} \thicksim \text{N}(\mu_{\beta_{j}}, \, \Sigma_{\beta_{j}})$ where
	\begin{equation*}
	\setlength{\jot}{10pt}
	\begin{aligned}
	\mu_{\beta_{j}} &= \Sigma_{\beta_{j}}(\sigma^{2})^{-1}\sum_{i=1}^{n}U_{ij}\tilde{y}_{ij} \\
	\Sigma_{\beta_{j}} &= \sigma^{2}\left(\sum_{i=1}^{n}U_{ij}U_{ij}^\top + V_{j}^{-1}\right)^{-1}
	\end{aligned}
	\end{equation*}
	and $\tilde{y}_{ij}$ is defined as $\tilde{y}_{ij}=y_{i}-W_{i}^\top\alpha-E_{i}^\top\theta-U_{(j)}^\top\beta_{(j)}$.
	
	\item $r_{j}^{-1}|\text{rest} \thicksim \text{Inv-Gaussian}(\eta_{1}, \, \sqrt{\frac{\eta_{1}\sigma^{2}}{\lVert\beta_{j}\rVert_{2}^{2}}})$
	
	\item $\omega_{jl}^{-1}|\text{rest} \thicksim \text{Inv-Gaussian}(\eta_{2}, \, \sqrt{\frac{\eta_{2}\sigma^{2}}{\beta_{jl}^{2}}})$
	
	\item $\eta_{1}|\text{rest} \thicksim \text{Gamma}\left(s_{\eta_{1}}, \, r_{\eta_{1}} \right)$, where $s_{\eta_{1}}=\frac{p}{2}+1$ and the rate parameter $r_{\eta_{1}}=\frac{\sum_{j=1}^{p}r_{j}}{2} + d_{1}$. 
	
	\item $\eta_{2}|\text{rest} \thicksim \text{Gamma}\left(s_{\eta_{2}}, \, r_{\eta_{2}} \right)$, where $s_{\eta_{2}}=p\times L+1$ and the rate parameter $r_{\eta_{2}}=\frac{\sum_{j,l}\omega_{j}}{2} + d_{2}$. 
	
	\item $\sigma^{2}|\text{rest} \thicksim \text{Inv-Gamma}(\frac{n+p\times L}{2},\; \frac{\sum_{i=1}^{n}\tilde{y_{i}}^{2}+\sum_{j=1}^{p}\beta_{j}^\top V_{j}^{-1}\beta_{j} }{2})$, where $\tilde{y_{i}}=y_{i}-W_{i}^\top\alpha-E_{i}^\top\theta-U_{i}^\top\beta$.
	
\end{itemize}

\subsection{BGL}
\subsubsection{Hierarchical model specification}

\setlength{\jot}{7pt}
\begin{gather*}
Y \propto (\sigma^{2})^{-\frac{n}{2}} \exp\left\{-\frac{1}{2\sigma^{2}} \sum_{i=1}^{n}(y_{i}- W_{i}^\top\alpha - E_{i}^\top\theta - U_{i}^\top\beta)^{2} \right\} \\ 
\alpha \thicksim \text{N}_{q}(0, \, \Sigma_{\alpha0}) \\
\theta \thicksim \text{N}_{k}(0, \, \Sigma_{\theta0}) \\
\beta_{j}|\sigma^{2}, s_{j} \stackrel{ind}{\thicksim} \text{N}_{L}\left(0,\, \sigma^{2}s_{j}\textbf{I}_{L}\right) \quad j=1,\dots,p\\
s_{j}|\eta \stackrel{ind}{\thicksim} \text{Gamma}\left(\frac{L+1}{2}, \frac{\eta}{2}\right) \quad j=1,\dots,p\\
\eta \thicksim \text{Gamma}\left(d_{1},\; d_{2}\right)\\
\sigma^{2} \thicksim 1/\sigma^{2}
\end{gather*}

\subsubsection{Gibbs Sampler}

\begin{itemize}
	
	\item $\alpha|\text{rest} \thicksim \text{N}(\mu_{\alpha}, \, \Sigma_{\alpha})$, where
	\begin{equation*}
	\setlength{\jot}{10pt}
	\begin{aligned}
	\mu_{\alpha} &= \Sigma_{\alpha}(\sigma^{2})^{-1}\sum_{i=1}^{n}W_{i}(y_{i}-E_{i}^\top\theta-U_{i}^\top\beta) \\
	\Sigma_{\alpha} &= \left(\frac{1}{\sigma^{2}}\sum_{i=1}^{n}W_{i}W_{i}^\top + \Sigma_{\alpha0}^{-1} \right)^{-1} 
	\end{aligned}
	\end{equation*}
	
	\item $\theta|\text{rest} \thicksim \text{N}(\mu_{\theta}, \, \Sigma_{\theta})$, where
	\begin{equation*}
	\setlength{\jot}{10pt}
	\begin{aligned}
	\mu_{\theta} &= \Sigma_{\theta}(\sigma^{2})^{-1}\sum_{i=1}^{n}E_{i}(y_{i}-W_{i}^\top\alpha-U_{i}^\top\beta) \\
	\Sigma_{\theta} &= \left(\frac{1}{\sigma^{2}}\sum_{i=1}^{n}E_{i}E_{i}^\top + \Sigma_{\theta0}^{-1} \right)^{-1} 
	\end{aligned}
	\end{equation*}
	
	\item $\beta_{j}|\text{rest} \thicksim \text{N}(\mu_{\beta_{j}}, \, \sigma^{2}\Sigma_{\beta_{j}})$ where
	\begin{equation*}
	\setlength{\jot}{10pt}
	\begin{aligned}
	\mu_{\beta_{j}} &= \Sigma_{\beta_{j}}\sum_{i=1}^{n}U_{ij}\tilde{y}_{ij} \\
	\Sigma_{\beta_{j}} &= \left(\sum_{i=1}^{n}U_{ij}U_{ij}^\top + \frac{1}{s_{j}}\textbf{I}_{L} \right)^{-1}
	\end{aligned}
	\end{equation*}
	and $\tilde{y}_{ij}$ is defined as $\tilde{y}_{ij}=y_{i}-W_{i}^\top\alpha-E_{i}^\top\theta-U_{(j)}^\top\beta_{(j)}$.
	
	\item $s_{j}^{-1}|\text{rest} \thicksim \text{Inverse-Gaussian}(\eta, \, \sqrt{\frac{\eta \sigma^{2}}{\lVert\beta_{j}\rVert_{2}^{2}}})$
	
	
	\item $\eta|\text{rest} \thicksim \text{Gamma}\left(s_{\eta}, \, r_{\eta} \right)$, where $s_{\eta}=\frac{p+p\times L}{2}+d_{1}$ and the rate parameter $r_{\eta}=\frac{\sum_{j=1}^{p}s_{j}}{2} + d_{2}$. 
	
	\item $\sigma^{2}|\text{rest} \thicksim \text{Inv-Gamma}(\frac{n+p\times L}{2} ,\; \frac{\sum_{i=1}^{n}\tilde{y_{i}}^{2} + \sum_{j=1}^{p}(s_{j})^{-1}\beta_{j}^\top \beta_{j} }{2})$, where $\tilde{y_{i}}=y_{i}-W_{i}^\top\alpha-E_{i}^\top\theta-U_{i}^\top\beta$.
	
\end{itemize}

\subsection{BL}
\subsubsection{Hierarchical model specification}

\setlength{\jot}{7pt}
\begin{gather*}
Y \propto (\sigma^{2})^{-\frac{n}{2}} \exp\left\{-\frac{1}{2\sigma^{2}} \sum_{i=1}^{n}(y_{i}- W_{i}^\top\alpha - E_{i}^\top\theta - U_{i}^\top\beta)^{2} \right\} \\ 
\alpha \thicksim \text{N}_{q}(0, \, \Sigma_{\alpha0}) \\
\theta \thicksim \text{N}_{k}(0, \, \Sigma_{\theta0}) \\
\beta_{jl}|\sigma^{2}, s_{jl} \stackrel{ind}{\thicksim} \text{N} \left(0,\, \sigma^{2} s_{jl}\right) \quad j=1,\dots,p;\; l=1,\dots,L \\
s_{jl}|\eta \stackrel{ind}{\thicksim} \text{Gamma}\left(1, \frac{\eta}{2}\right) \quad j=1,\dots,p;\;  l=1,\dots,L\\
\eta \thicksim \text{Gamma}\left(d_{1},\; d_{2}\right)\\
\sigma^{2} \thicksim 1/\sigma^{2}
\end{gather*}

\subsubsection{Gibbs Sampler}

\begin{itemize}
	
	\item $\alpha|\text{rest} \thicksim \text{N}(\mu_{\alpha}, \, \Sigma_{\alpha})$, where
	\begin{equation*}
	\setlength{\jot}{10pt}
	\begin{aligned}
	\mu_{\alpha} &= \Sigma_{\alpha}(\sigma^{2})^{-1}\sum_{i=1}^{n}W_{i}(y_{i}-E_{i}^\top\theta-U_{i}^\top\beta) \\
	\Sigma_{\alpha} &= \left(\frac{1}{\sigma^{2}}\sum_{i=1}^{n}W_{i}W_{i}^\top + \Sigma_{\alpha0}^{-1} \right)^{-1} 
	\end{aligned}
	\end{equation*}
	
	\item $\theta|\text{rest} \thicksim \text{N}(\mu_{\theta}, \, \Sigma_{\theta})$, where
	\begin{equation*}
	\setlength{\jot}{10pt}
	\begin{aligned}
	\mu_{\theta} &= \Sigma_{\theta}(\sigma^{2})^{-1}\sum_{i=1}^{n}E_{i}(y_{i}-W_{i}^\top\alpha-U_{i}^\top\beta) \\
	\Sigma_{\theta} &= \left(\frac{1}{\sigma^{2}}\sum_{i=1}^{n}E_{i}E_{i}^\top + \Sigma_{\theta0}^{-1} \right)^{-1} 
	\end{aligned}
	\end{equation*}
	
	\item $\beta_{jl}|\text{rest} \thicksim \text{N}(\mu_{\beta_{jl}}, \, \sigma^{2}_{\beta_{jl}})$ where
	\begin{equation*}
	\setlength{\jot}{10pt}
	\begin{aligned}
	\mu_{\beta_{jl}} &= \sigma^{2}_{\beta_{jl}}(\sigma^{2})^{-1}\sum_{i=1}^{n}U_{ijl}\tilde{y}_{ijl} \\
	\sigma^{2}_{\beta_{jl}} &= \sigma^{2}\left(\sum_{i=1}^{n}U_{ijl}^{2} + \frac{1}{s_{jl}} \right)^{-1}
	\end{aligned}
	\end{equation*}
	and $\tilde{y}_{ijl}$ is defined as $\tilde{y}_{ijl}=y_{i}-W_{i}^\top\alpha-E_{i}^\top\theta-U_{(jl)}^\top\beta_{(jl)}$.

	\item $s_{j}^{-1}|\text{rest} \thicksim \text{Inverse-Gaussian}(\eta, \, \sqrt{\frac{\eta \sigma^{2}}{\beta_{jl}^{2}}})$

	
	\item $\eta|\text{rest} \thicksim \text{Gamma}\left(s_{\eta}, \, r_{\eta} \right)$, where $s_{\eta}=p\times L+d_{1}$ and the rate parameter $r_{\eta}=\frac{\sum_{j,l}s_{jl}}{2} + d_{2}$. 
	
	\item $\sigma^{2}|\text{rest} \thicksim \text{Inv-Gamma}(\frac{n+p\times L}{2},\; \frac{\sum_{i=1}^{n}\tilde{y_{i}}^{2}+\sum_{j,l}s_{jl}^{-1}\beta_{jl}^{2}}{2})$, where $\tilde{y_{i}}=y_{i}-W_{i}^\top\alpha-E_{i}^\top\theta-U_{i}^\top\beta$.
	
\end{itemize}

\end{document}